%% file: main.tex
    \documentclass[a4,12pt]{article}
  
    \usepackage{abstract} 
    
    \usepackage{titlesec} 

    \usepackage{authblk} 

	\usepackage{datetime} 
    \newdateformat{usvardate}{
  	\monthname[\THEMONTH] \ordinal{DAY}, \THEYEAR}

  	\usepackage[bottom]{footmisc} 
   
    \usepackage{fancyhdr}
    \fancyheadoffset{0cm}
    \providecommand{\keywords}[1]{\textbf{\textit{Keywords ---}} #1}

\newcommand\wordcount{%
  \immediate\write18{texcount -utf8 -merge -sum -incbib -dir -sub=none -brief \jobname.tex | cut -d : -f 1 > 'count.txt'}%
  \input{count.txt}\ignorespaces words%
}

    \usepackage{amssymb}
    \usepackage{amsthm}
    \usepackage{newtxmath}
    \usepackage{mathtools}

    \usepackage{microtype} 
    \usepackage[utf8]{inputenc}
    \usepackage{newtxtext} 
    \usepackage{xcolor}
    
	\usepackage{setspace} 
    \usepackage{lineno,xcolor} 	
    \usepackage{xurl}
    \usepackage{breakcites}
	\usepackage[top=1.5in, bottom=1.5in, left=1in, right=1in]{geometry}

	\usepackage[colorinlistoftodos]{todonotes} 
	\usepackage{soul} 
    
    \usepackage{graphicx} 
    \usepackage{float} 

    \usepackage{printlen}

    \usepackage{multirow}
    \usepackage{caption} 
    \usepackage{longtable} 
    \usepackage{tabularx} 
    \newcolumntype{L}[1]{>{\raggedright\arraybackslash}p{#1}}
    \newcolumntype{C}[1]{>{\centering\arraybackslash}p{#1}}
    \newcolumntype{R}[1]{>{\raggedleft\arraybackslash}p{#1}}
    \usepackage{relsize} 
     \usepackage{arydshln}
     \usepackage{adjustbox} 
     \usepackage{tabularray}
     \usepackage[flushleft]{threeparttable}

    \usepackage{algorithm}
    \usepackage{algpseudocode}

    \usepackage{hyperref}
    \usepackage[
    backend=bibtex,
    style=nature,
    ]{biblatex}
\begin{document}


\input{coverpage} 

\thispagestyle{empty} 

\input{title_noauthors}

\maketitle 


\begin{abstract}
\noindent A central characteristic of Bayesian statistics is the ability to consistently incorporate prior knowledge into various modeling processes. In this paper, we focus on translating domain expert knowledge into corresponding prior distributions over model parameters, a process known as prior elicitation. Expert knowledge can manifest itself in diverse formats, including information about raw data, summary statistics, or model parameters. A major challenge for existing elicitation methods is how to effectively utilize all of these different formats in order to formulate prior distributions that align with the expert's expectations, regardless of the model structure. 
To address these challenges, we develop a simulation-based elicitation method that can learn the hyperparameters of potentially any parametric prior distribution from a wide spectrum of expert knowledge using stochastic gradient descent.
We validate the effectiveness and robustness of our elicitation method in four representative case studies covering linear models, generalized linear models, and hierarchical models.
Our results support the claim that our method is largely independent of the underlying model structure and adaptable to various elicitation techniques, including quantile-based, moment-based, and histogram-based methods. 
\end{abstract}

\setlength\parindent{.45in} \keywords{Bayesian statistics, prior specification, prior elicitation, simulation-based inference, expert knowledge}


\section{Introduction}
The essence of Bayesian statistics lies in the ability to consistently incorporate prior knowledge into the modeling process \supercite{jaynes2003probability, gelman2013bayesian}. Accordingly, specifying sensible prior distributions over the parameters of Bayesian models can have multiple advantages. For instance, integrating theory-guided knowledge into otherwise data-driven models effectively constrains the model behavior within an expected range \supercite{navarro2019between, mikkola2023prior, manderson2023translating}. This integration ideally enhances both model faithfulness and computational aspects, including convergence and sampling efficiency. Moreover, well-specified prior distributions can positively influence various model criteria, such as parameter recoverability, predictive performance, and convergence \supercite{burkner2022some}.

However, despite these apparent advantages, prior specification poses a significant challenge for analysts as it is often unclear a priori what constitutes a ``sensible'' prior \supercite{gelman2017prior}. In this paper, we focus on one specific approach to prior specification, namely, the elicitation and translation of expert knowledge into prior distributions, also known as \emph{prior elicitation} \supercite{mikkola2023prior}. Against this background, \textit{a sensible prior is one that accurately reflects domain knowledge as elicited from an expert or a group of experts}. Still, achieving the latter criterion poses challenges on its own. Model parameters for which priors are needed might lack intuitive meaning for the domain expert, making it difficult to query the properties of these parameters directly \supercite{albert2012combining}. Moreover, the relationship between priors and the data may not be apparent from the model specification, especially for complex models, such as hierarchical models \supercite{da2019prior}. Indeed, constructing priors for every single model parameter in models with a large number of parameters might be inefficient or even infeasible. Finally, the domain expert might lack statistical expertise to translate their knowledge into appropriate prior distributions \supercite{hartmann2020flexible}.

To address these challenges, several tools for prior elicitation have been developed in the past, as reviewed by Garthwaite et al. \supercite{garthwaite2005statistical} and  O'Hagan et al.\supercite{o2006uncertain}. However, despite the widespread acceptance and routine application of Bayesian statistics nowadays, the field of prior elicitation still lags behind in terms of its routine implementation by practitioners (see Mikkola et al.\supercite{mikkola2023prior} for a recent comprehensive review). One contributing factor is that many existing methods primarily aim to elicit information about the model parameters directly. This approach makes these methods inherently model-specific and limits their widespread applicability. Additionally, as mentioned previously, the emphasis on parameters can present a challenge for experts in terms of interpretability \supercite{da2019prior,bedrick1996new,kadane1980interactive}.  

In recent years, there has been an increasing focus on the development of model-agnostic approaches that center around the prior predictive distribution \supercite{manderson2023translating}. These methods allow for the integration of expert knowledge regarding observed data patterns (i.e., elicitation in the observable space). In contrast to interpreting model parameters, domain experts can usually effectively interpret the scale and magnitude of observable quantities \supercite{mikkola2023prior,muandet2017kernel,akbarov2009probability,da2019prior,hartmann2020flexible}. Despite these recent developments, the general applicability as well as the actual application of elicitation methods remain limited \supercite{mikkola2023prior}. This lack of popularity persists, at least in part, because these methods are relatively complex, do not easily generalize to different types of expert information, or necessitate substantial tuning or other manual adjustments. In light of the preceding considerations, we introduce an elicitation method that seeks to overcome these challenges. Specifically, this work makes a contribution to prior elicitation research by proposing a method that satisfies the following criteria:

\begin{enumerate}
    \item \emph{Model Independence}: Our method is agnostic to the specific probabilistic model, as long as sampling from it is feasible and stochastic gradients can be computed.
    \item \emph{Effective Utilization of Expert Knowledge}: By incorporating diverse expert information on model parameters, observed data patterns, or other relevant statistics, our method maximizes the utility of expert knowledge and expands the diversity of information embedded in the  model.
    \item \emph{Flexibility in Knowledge Formats and Elicitation Techniques}: Our method can accommodate various knowledge formats and adapt to different elicitation techniques, ensuring that individual expert preferences are considered.
    \item \emph{Modular Design}: Having a modular structure, our method provides flexibility to analysts, allowing for easy adaptation, improvement, or replacement of specific components, both during method development and application.
\end{enumerate}

\section{Related Work}\label{sec:related-work}
The process of prior elicitation involves the extraction and translation of knowledge from domain experts with the goal to specify appropriate prior distributions for the parameters in probabilistic models. For a comprehensive overview of this field, we refer interested readers to the recent review provided by Mikkola et al.\supercite{mikkola2023prior}. In prior elicitation, a distinction can be made between techniques for knowledge \emph{extraction} \supercite{garthwaite2005statistical,o2006uncertain,falconer2022methods}, and methods for \emph{translating} knowledge into suitable prior distributions. We explicitly aim to contribute to the second. 

Historically, elicitation methods have primarily focused on prior elicitation in a model's parameter space, entailing direct inquiries to experts regarding the model parameters \supercite{mikkola2023prior}. However, there has been a recent shift towards methods that facilitate elicitation in the observable space \supercite{manderson2023translating}. The current vision of a (yet to be achieved) ``gold standard'' is an elicitation method that includes both a model's parameter and observable space, exhibits model-agnostic characteristics, and prioritizes sample efficiency to minimize the human effort involved \supercite{mikkola2023prior}. Taking these desiderata into consideration, our method builds upon recent advancements in prior elicitation, specifically on the works of da Silva et al.\supercite{da2019prior}, Hartmann et al.\supercite{hartmann2020flexible}, and Manderson \& Goudie\supercite{manderson2023translating} who proposed model-agnostic elicitation methods. 
All three methods have in common that they learn hyperparameters of prior distributions by minimizing the discrepancy between model-implied and expert-elicited statistics. These quantities are associated with the observable space, leading to the model-agnostic characteristic of these approaches. Differences among these methods arise from the specification of target quantities, discrepancy measures, and the specific optimization procedure.

Manderson \& Goudie\supercite{manderson2023translating} adopt a two-stage global optimization process that incorporates multi-objective Bayesian optimization, while our approach aligns more with the methods proposed by da Silva et al.\supercite{da2019prior} and Hartmann et al.\supercite{hartmann2020flexible}, who employ stochastic gradient-based optimization.
However, similarly to our approach, Manderson \& Goudie\supercite{manderson2023translating} assume that an expert is queried about observable or model-derived quantities at various quantiles. 
Their method then uses these elicited statistics to fit a parametric distribution whose cumulative density function becomes the input to the discrepancy measure. 
In contrast, our method imposes no constraints on the domain of the discrepancy measure; quantiles, histograms, or moments can all serve as inputs.

The utilization of quantile-based elicitation can also be found in Hartmann et al.\supercite{hartmann2020flexible}, where the authors directly model the probability of the outcome data using a Dirichlet process. However, the work by da Silva et al.\supercite{da2019prior} is most closely related to our method, as the authors propose a generic methodology that in principal supports an arbitrary choice of target quantities based on the prior predictive distribution and discrepancy measures. They illustrate this by selecting the expected mean and variance as inputs for a discrepancy measure that simultaneously optimizes both. However, it is not apparent how the method can be readily applied to arbitrary target quantities without the need for considerable customization. In contrast, due to the generality of the MMD criterion, our method is broadly applicable to various target quantities without the need for extensive customization. Furthermore, it differs from previous methods by allowing the elicited information to refer to both parameters \emph{and} observables or other model-derived quantities. This flexibility is achieved by treating the model-implied target quantities in their most general form as a function of the model parameters.

Finally, an essential feature of our method is the use of simulations to obtain prior hyperparameter inference, which classifies it as a variant of \emph{simulation-based inference} (SBI), also known as likelihood-free inference \supercite{cranmer2020frontier}. SBI circumvents the analytic intractability of complex models through the use of various simulation-based techniques (e.g., training a deep neural network to recover model parameters from data). Analogously, our method utilizes simulations to approximate the potentially intractable distribution of the target quantities given the hyperparameters.

\input{method_section}

\section{Case Studies}\label{sec:case-studies}
Below, we present four case studies exemplifying the application of our method. Each study concentrates on a distinct statistical model, highlighting specific facets of our approach.
To introduce the method, in Case Study 1 (Section \ref{subsec: case-study1}) we adopt a classic approach: a normal linear regression model with two factors. Proceeding to Case Study 2 (Section \ref{subsec: case-study2}), we demonstrate the performance of our method for discrete distributions using a binomial distribution with a logit link. Given that the Softmax-Gumbel trick necessitates a double-bounded distribution, we probe the application of our method to a Poisson distribution with varying upper truncation thresholds in Case Study 3 (Section \ref{subsec: case-study3}).  
Finally, in Case Study 4 (Section \ref{subsec: case-study4}), we employ our method within hierarchical models, learning prior distributions for both overall and group-specific (varying) effects under multiple likelihood assumptions as well as partially inconsistent expert information.
For an overview of the employed models and the core motivations driving each case study, consult Table \ref{tab:overview-case-studies}.
\begin{table}[ht]
    \caption{\textit{Overview Case Studies.}}
    \centering
    \begin{tabular}{p{0.12\textwidth} p{0.12\textwidth} p{0.61\textwidth}}
    \hline
    \emph{Case Study} & \emph{Model} & \emph{Core motivation}         \\
    \hline
     Section \ref{subsec: case-study1}  & Normal & Introduce method         \\
     Section \ref{subsec: case-study2}           & Binomial & Double-bounded discrete RV \\
     Section \ref{subsec: case-study3}           & Poisson  & Lower-bounded discrete RV with upper truncation threshold \\
    Section \ref{subsec: case-study4}          & Hierarchical &  Normal model including varying effects \\
    \hline
    \end{tabular}
    \label{tab:overview-case-studies}
\end{table}
All code and material can be found on GitHub \url{https://github.com/florence-bockting/PriorLearning}, our project website \url{https://florence-bockting.github.io/PriorLearning/index.html}, and our online supplement \url{https://osf.io/rxgv2}.

\subsection{General Setup}
\paragraph{Learning Algorithm}
In each case study, we utilize mini-batch SGD to learn all model hyperparameters. Each optimization process is characterized by a set of \emph{algorithm parameters} including the batch size ($B$), the number of epochs ($E$), the number of samples from the prior distributions ($S$), and the initial as well as minimum learning rates ($\phi^0$ and $\phi^\text{min}$) of the
exponential decay learning rate schedule used with the Adam optimizer. The specific settings of the optimization process are fully described in the respective sections. All case studies were implemented in Python, utilizing the \textit{TensorFlow} library \supercite{abadi2016tensorflow}, and optimization was performed on a CPU-only machine. 

\paragraph{Ideal Expert Model}
In the case studies presented here, we employ an expert model to simulate responses, rather than querying real experts. This choice is motivated by our aim to demonstrate the \emph{validity} of our method. By \emph{validity}, we refer to our method's ability in accurately recovering the ``true'' hyperparameter values. This objective presupposes our knowledge of these true hyperparameter values, which is possible by the use of expert simulations. An important future step will involve assessing our method with actual experts.

In each case study, we introduce a specific data-generating model corresponding to an exemplary application scenario. The expert model mirrors this data-generating model precisely, with a distinct set of hyperparameter values $\lambda^*$ that fully defines the statistical model. Consequently, we assume that the expert has exact knowledge of the generative model, which is why we refer to it as the ``ideal'' expert. The specific values of $\lambda^*$ are introduced in each case study. From this ideal expert model, we simulate all predefined target quantities (discussed next) and compute the corresponding elicited statistics. For instance, we simulate a distribution of group means (target quantity) and derive the corresponding quantiles (elicited statistic using quantile-based elicitation). These elicited quantiles serve as input, representing the ``expert knowledge'', for our method to learn the hyperparameters $\lambda$. 
The expert data is simulated only \emph{once before} the learning process, reflecting a one-shot elicitation scenario. The use of the ideal expert approach enables us to assess the validity of our method by evaluating its ability to recover the true hyperparameter values~$\lambda^*$. 

\paragraph{Selection of Target Quantities \& Elicited Statistics}
In the selection of target quantities and elicited statistics, we consider several key aspects. First, we aim to represent the variability of target quantities, which includes diverse statistics (e.g., $R^2$), predictions for the outcome variable (e.g., predictions for specific design points), and information about model parameters. Second, we account for the variability of elicitation techniques by employing histogram, quantile-based, and moment-based elicitation to compute the elicited statistics. Third, we prioritize interpretability by choosing target quantities that can be readily understood by experts in realistic scenarios. Finally, the selected number of elicited statistics should guarantee model identification. Addressing the issue of identification is crucial for our objective of assessing the validity of our method through the successful recovery of the true hyperparameters $\lambda^*$. To achieve this, the model must possess sufficient relevant information to uniquely identify the hyperparameter values. Otherwise, a failure to recover the hyperparameters could be attributed to either methodological limitations or insufficient information available from the elicited statistics.
In our case studies, we strive to achieve perfect model identification by carefully selecting a ``sufficiently large'' set of queried quantities. Specifically, in the quantile-based elicitation, we query the expert with respect to nine quantiles $Q_p$ with $p=(0.1, 0.2, \ldots, 0.9)$. For moment-based elicitation, the mean and standard deviation of the target
quantity, and for histogram elicitation, a histogram comprising $S$ observations is queried from the (ideal) expert (further specifications can be found in each case study).

We acknowledge that in practical applications of elicitation methods, a trade-off arises between the desired information sought from the expert to achieve model identification and the limitations imposed by the expert's available resources and knowledge. Since the goal of the case studies is to establish the validity of our method, our focus lies on achieving \textit{complete} prior hyperparameter identification. In Section \ref{sec:discussion-conclusion}, we will discuss future directions and steps, including the exploration of settings where prior hyperparameters are not fully identified by the provided expert information.

\subsection{Case Study 1: Normal Linear Regression} 
\label{subsec: case-study1}
\paragraph{Setup}
In the first case study, we present our method using a normal linear regression model, along with an example from the field of social cognition. Specifically, we focus on an experiment conducted by Unkelbach \& Rom\supercite{unkelbach2017referential}, where participants are presented with general knowledge statements in two consecutive phases. During the second phase, they are required to indicate whether each statement is true or false. The response variable of interest is the proportion of true judgments (PTJs). The main objective of the study is to investigate the influence of two factors on PTJs: (1) repetition (ReP), which involves presenting some statements from the first phase again in the second phase, and (2) encoding depth (EnC), where participants are divided into groups varying in the level of elaboration required for the sentences in the first phase. We employ a 2 (ReP: \emph{repeated, new}) $\times$ 3 (EnC: \emph{shallow, standard, deep}) between-subject factorial design with treatment contrasts used for both factors. The baseline levels are set as new for ReP and shallow for EnC. Following Unkelbach \& Rom\supercite{unkelbach2017referential}, we employ a linear model to describe the data-generating process 
\begin{align}\label{eq:linear-regression_model}
\begin{split}
    y_i &\sim \textrm{Normal}(\theta_i, s) \\
    \theta_i &= \beta_0 + \beta_1x_1 + \beta_2x_2 + \beta_3x_3 + \beta_4x_4 + \beta_5x_5\\
    \beta_k &\sim \textrm{Normal}(\mu_k, \sigma_k) \quad \textrm{ for }k=0,\ldots,5\\
    s &\sim \textrm{Exponential}(\nu).\\
\end{split}
\end{align}
The responses $y_i$ for each observation $i=1, \ldots, N$ are normally distributed with mean $\theta_i$ and standard deviation $s$. The expected value $\theta_i$ is modeled as a linear function of ReP and EnC. The regression coefficients $\beta_k$ for $k=0,\ldots,5$ are assigned normal prior distributions. Specifically, $\beta_0$ represents the PTJs for new statements in the shallow-encoding condition, while $\beta_1$ represents the difference in PTJs ($\Delta$PTJ) between repeated and new statements. Additionally, $\beta_2$ captures the $\Delta$PTJ for new statements in the shallow- vs. standard-encoding condition, $\beta_3$ represents the $\Delta$PTJ for new statements in the shallow- vs. deep-encoding condition, and $\beta_4$ and $\beta_5$ account for the interaction effects between ReP and EnC. The standard deviation $s$ of the normal likelihood follows an Exponential prior distribution with rate parameter $\nu$.
In our implementation, we employ a reparameterization trick for $s \sim \textrm{Exponential}(\nu)$, achieved by expressing a prior on the mean of $N$ replicated $s$ as $1/N \sum_n s_n \sim \textrm{Gamma}(N, N \cdot \nu)$. Our simulation results demonstrate improved learning behavior for this alternative formulation.
The goal is to learn a total of 13 hyperparameters, $\lambda = (\mu_k, \sigma_k, \nu)$, through expert knowledge.

\begin{figure*}
    \centering
    \includegraphics[width=\textwidth]{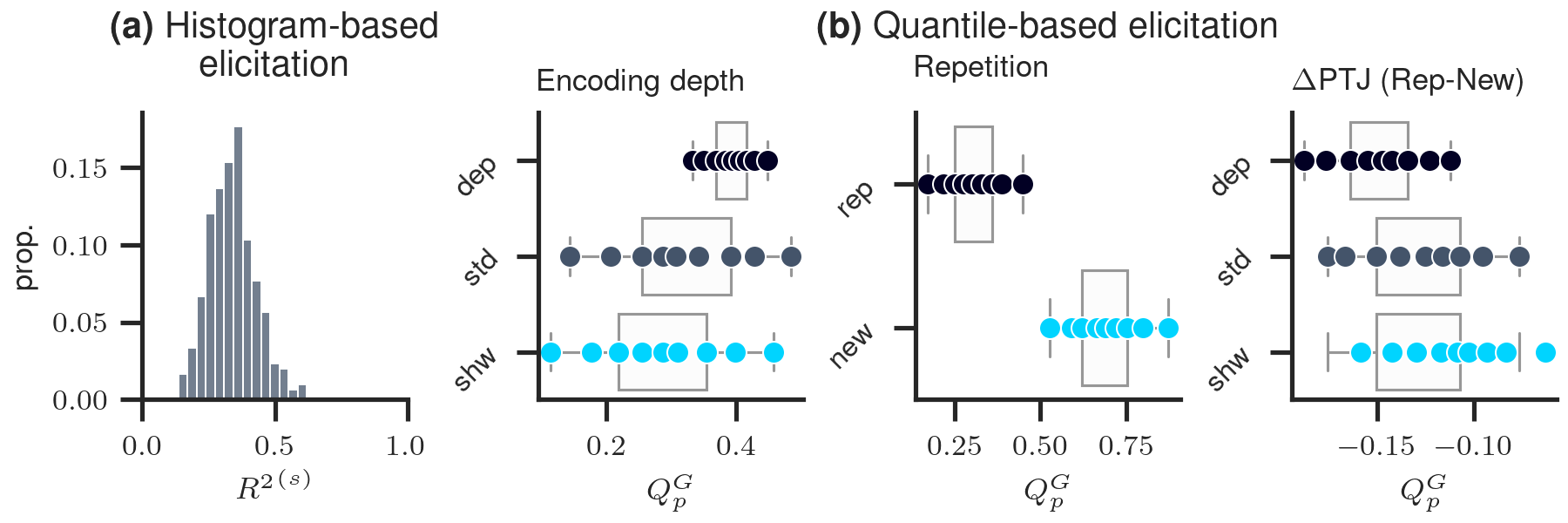}
    \caption{\emph{Expert-elicited statistics.} \textbf{(a)} elicited histogram of $R^2$ from the ideal expert; \textbf{(b)} nine expert-elicited quantiles for each remaining target quantity (see text for detailed information). Abbreviations: For the factor \emph{Encoding depth}: dep-deep, std-standard, and shw-shallow and for the factor \emph{Repetition}: rep-repeated and new.}
    \label{fig:cs1-target-quant}
\end{figure*}

\paragraph{Elicitation procedure}
We assume that the analyst elicits four target quantities from the expert: the expected PTJ for the marginal distribution of both factors EnC (1) and ReP (2), the expected $\Delta$PTJ between repeated and new statements for each EnC level (3), and the expected $R^2$ defined as defined as a variance ratio of the modeled predictive means and the predictive observations including the residual variance, $R^2= \textrm{var}(\theta_i) / \textrm{var}(y_i)$ (4).
e assume this information is queried from the expert
using quantile elicitation for (1-3) and histogram elicitation for (4). To model the ideal expert, we assume the following ``true'' hyperparameters $\lambda^* = (\mu_0=0.12, \sigma_0=0.02, \mu_1=0.15, \sigma_1=0.02, \mu_2=-0.02, \sigma_2=0.06, \mu_3=-0.03, \sigma_3=0.06, \mu_4=-0.02, \sigma_4=0.03, \mu_5=-0.04, \sigma_5=0.03, \nu=9.00)$. The corresponding elicited statistics of the ideal expert are depicted in Figure~\ref{fig:cs1-target-quant}. The first column depicts the elicited histogram for $R^2$ and the remaining columns the results of quantile-based elicitation.

\paragraph{Optimization}
To instantiate the optimization process the hyperparameters $\lambda$ are randomly initialized as follows: $\mu_k \sim \text{Normal}(0, 0.1)$, $\log\sigma_k \sim \text{Uniform}(-2, -4)$, and $\log\nu \sim \text{Uniform}(1, 2)$, whereby the scale and rate parameter are initialized on the log scale. Subsequently, we simulate from the forward model and compute the corresponding model-implied target quantities, along with the elicited statistics. The discrepancy between the model-implied and expert-elicited statistics can then be computed and the hyperparameter updated. The learning process is considered completed once the maximum number of epochs has been reached. Details about the optimization algorithm can be found in Section~\ref{sec:methods} and the corresponding specification of the algorithm parameters can be found in Appendix~\ref{app:diagnostics-linreg}.
\begin{figure*}[ht]
    \centering
    \includegraphics{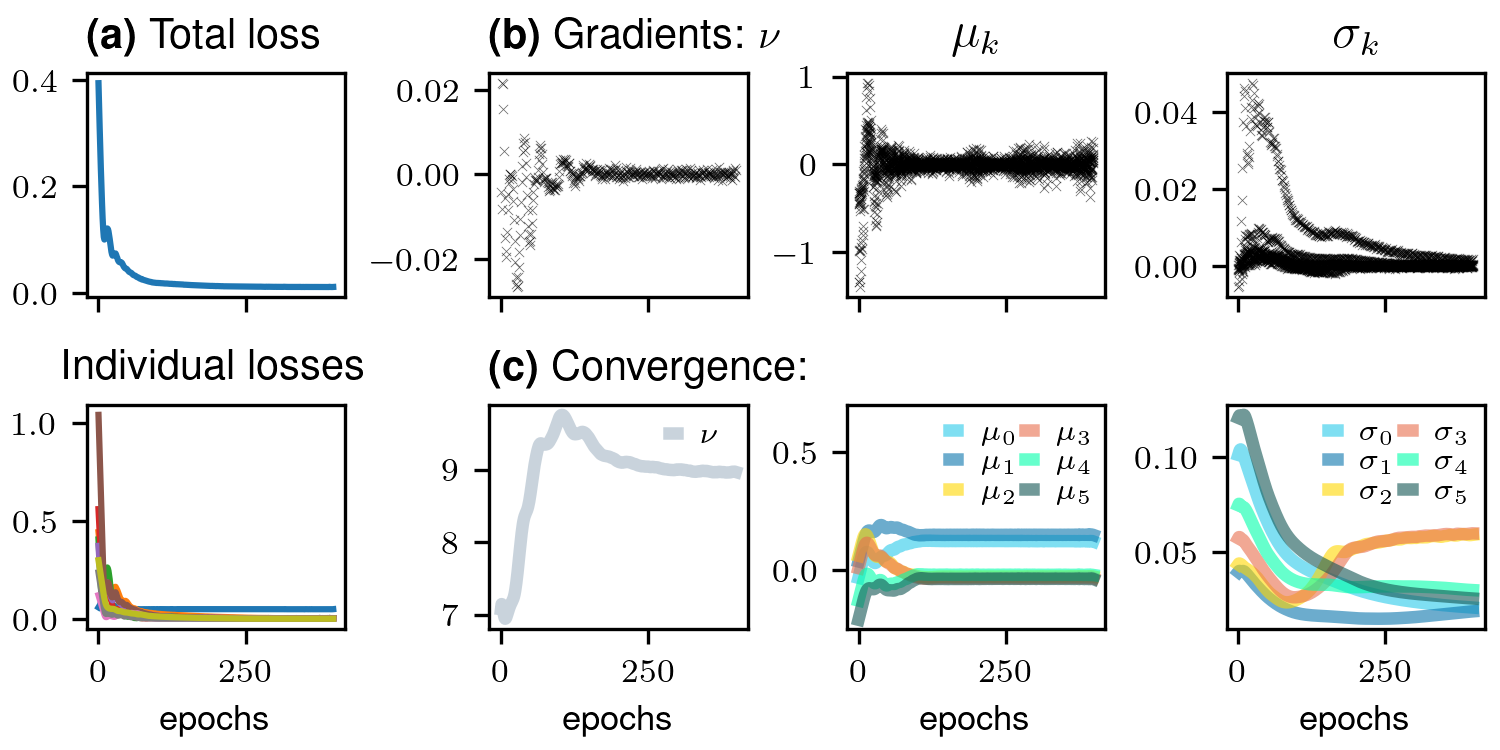}
    \caption{\emph{Convergence diagnostics for Case Study 1.} \textbf{(a)} loss value across epochs, demonstrating the desired decreasing trend of all loss values (i.e., the total loss and the individual loss components); \textbf{(b)} the expected decreasing trend towards zero of the gradients for each learned hyperparameter $\lambda$ is depicted; \textbf{(c)} update values of each learned hyperparameter after each iteration step (epoch), stabilizing in the long run at a specific value.}
    \label{fig:conv-diagn-linreg}
\end{figure*}
To assess whether learning was successful, we first check the \emph{convergence diagnostics} as summarized in Figure~\ref{fig:conv-diagn-linreg}.
Examining the loss functions depicted in the leftmost column demonstrates the desired decreasing behavior for both the total loss as well as the individual loss components. The gradients of the hyperparameters $\lambda$ are depicted in the upper, right row, indicating the expected decreasing behavior towards zero across time. Finally, convergence of hyperparameters $\lambda$ during the learning process is illustrated in the lower, right row. 

\begin{figure*}
    \centering
    \includegraphics{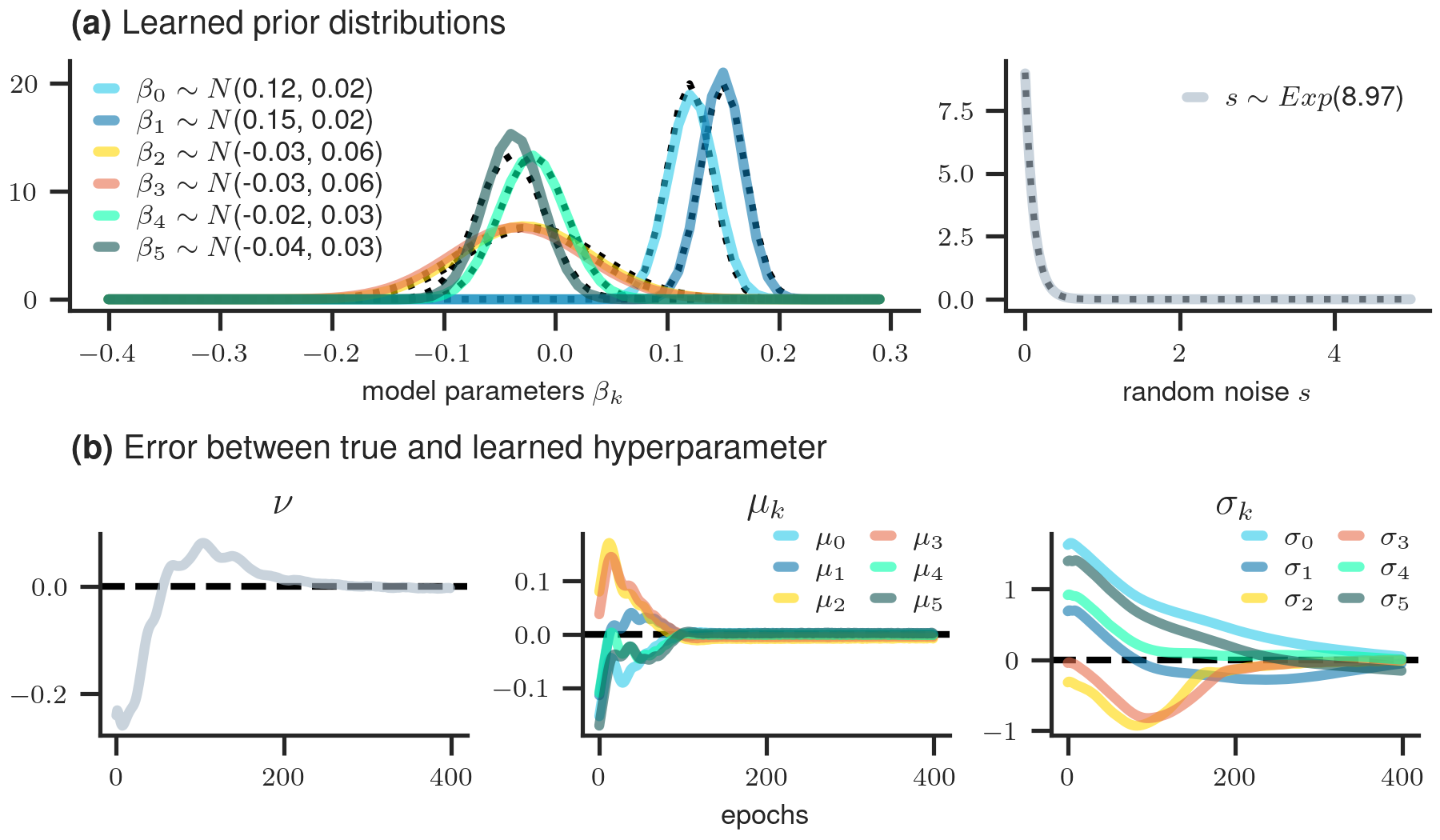}
    \caption{\emph{Results of Case Study 1.} \textbf{(a)} true (dotted line) and learned (solid line) prior distributions per model parameter $\beta_k$ and $s$; \textbf{(b)} error between learned and true hyperparameter values $(\nu, \mu_k, \sigma_k)$ over time.}
    \label{fig:linreg-sum-res}
\end{figure*}

\paragraph{Results}
After having confirmed successful convergence, we shift our focus to the simulation results as depicted in Figure \ref{fig:linreg-sum-res}. The final learned hyperparameter $\lambda$ is computed as the average of the last 30 epochs. The resulting \emph{learned} prior distributions are shown in the upper row of Figure \ref{fig:linreg-sum-res}. Solid lines indicate the learned priors and dotted lines the \emph{true} priors (according to the ideal expert). 
The substantial overlap between these distributions indicates a successful learning process. This is further emphasized in the second row, where the error between the learned and true hyperparameter values gradually decreases towards zero. 

The case study illustrates that our method effectively recovers true hyperparameters in a linear model, which is a special case within the broader framework of \emph{generalized linear models} (GLMs). In GLMs, a non-linear function can be used to transform the expected value, accounting for natural range restrictions in parameters like probabilities or counts. In the upcoming case studies we further explore the method's performance across various response distributions and link functions.

\subsection{Case Study 2: GLMs -- Binomial Model}
\label{subsec: case-study2}
\paragraph{Setup}
In Case Study 2 we utilize a Binomial response distribution with a logit-link function for the probability parameter. As an accompanying example, we use the Haberman's survival dataset from the UCI machine learning repository\supercite{Dua2019}. The dataset contains cases from a study that was conducted between 1958 and 1970 at the University of Chicago's Billings Hospital on the survival of patients who had undergone surgery for breast cancer. In the following, we use the detected number of axillary lymph nodes that contain cancer (i.e., (positive) axillary nodes) as numerical predictor $X$ which consists in total of 31 observations ranging between $0$ and $59$ axillary nodes. The dependent variable $y$ is the number of patients who died within five years out of $T=100$ trials for each observation $i=1,\ldots,N$. We consider a simple Binomial regression model with one continuous predictor 
\begin{align}
    \begin{split}
        y_{i} &\sim \textrm{Binomial}(T, \theta_i)\\
        \textrm{logit}(\theta_i) &= \beta_0 + \beta_1 x_i \\
        \beta_k &\sim \textrm{Normal}(\mu_k, \sigma_k) \quad \textrm{for} \quad k = 0,1.
    \end{split}
\end{align}
The probability parameter $\theta_i$ is predicted by a continuous predictor $x$ with an intercept $\beta_0$ and slope $\beta_1$. We assume normal priors for the regression coefficients, with mean $\mu_k$ and standard deviation $\sigma_k$ for $k=0,1$. Through the logit-link function, the probability $\theta_i$ is mapped to the scale of the linear predictor. The objective is to learn the hyperparameters $\lambda_k=(\mu_k, \sigma_k)$ based on expert knowledge.

\paragraph{Elicitation procedure \& optimization}
The analyst selects as target quantities the expected number of patients who died within five years for different numbers of axillary nodes $x_i$, with $i = 0, 5, 10, 15, 20, 25, 30$. The expert is queried for each selected design point using a quantile-based elicitation procedure. We define the ideal expert by the following true hyperparameters $\lambda^*=(\mu_0=-0.51, \sigma_0=0.06, \mu_1=0.26, \sigma_1=0.04)$. The specification of the algorithm parameters for the optimization procedure can be found in Appendix \ref{app:diagnostics-binom}.
The convergence diagnostics check follows the same procedure as discussed for Case Study~1, and showed successful convergence (see Appendix \ref{fig:conv-diag-binom}). 

\begin{figure*}[ht]
    \centering
    \includegraphics{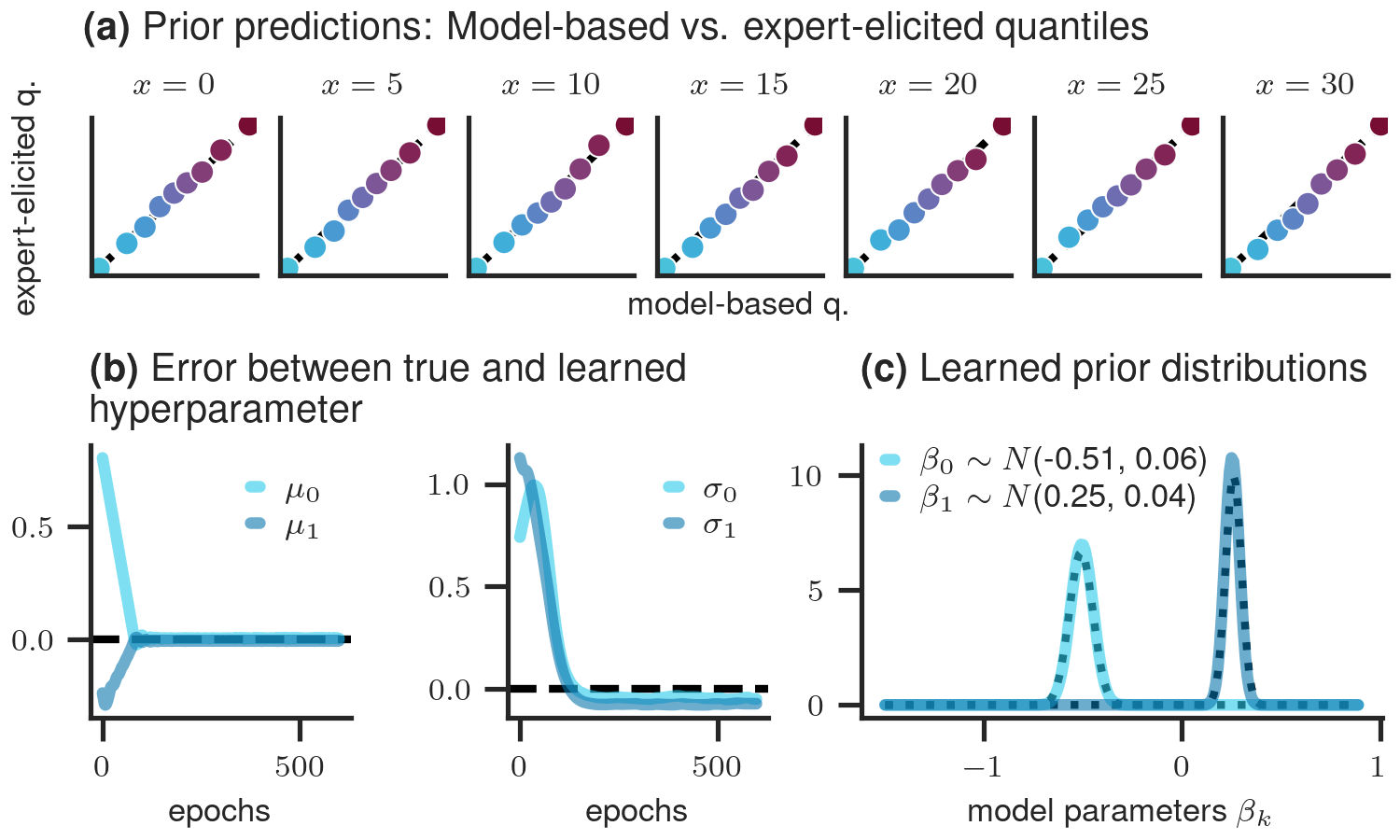}
    \caption{\emph{Results of Case Study 2}: \textbf{(a)} comparison between model-based and expert-elicited quantiles for each selected $x_i$; \textbf{(b)} learning of hyperparameters across epochs, showcasing the difference between the true and learned values; \textbf{(c)} true (dotted line) and learned (solid line) prior distributions of each model parameter.}
    \label{fig:binom-sum-res}
\end{figure*}

\paragraph{Results}
The simulation results, based on the final learned hyperparameters $\lambda$, are presented in Figure~\ref{fig:binom-sum-res}.
The upper row shows a comparison between the expert-elicited and model-based quantiles for each queried number of axillary nodes $x_i$, revealing an almost perfect match between both quantities. In the lower right panels, the error between the true and learned hyperparameters is depicted and indicates successful learning. Additionally, the lower right panel presents the true (dotted line) and learned (solid line) prior distributions which show a perfect match.

\subsection{Case Study 3: GLMs -- Poisson Model}
\label{subsec: case-study3}
\paragraph{Setup}
In Case Study~3, we extend our examination of count data likelihoods, with a specific focus on the Poisson distribution. Unlike the Binomial distribution, the Poisson distribution lacks an upper bound. This distinction becomes important, since we employ the Gumbel-Softmax trick during gradient-based learning of discrete random variables (see Section \ref{subsec:SGD} for details). The Gumbel-Softmax trick treats discrete distributions as categorical distributions with a finite number of categories, necessitating a double-bounded count distribution. To meet this requirement, we adopt the approach proposed by Joo et al.\supercite{joo2021generalized} and introduce a truncation threshold $t^u$ as the upper bound for the Poisson distribution. 

For demonstration purposes, we adapt an example from Johnson et al.\supercite{johnson2022bayes}, which  investigates the number of LGBTQ+ anti-discrimination laws in each US state. The distribution of these laws is assumed to follow a Poisson distribution, with the rate of such laws being influenced by demographic and voting trend. The demographic trend is quantified by the percentage of a state's residents living in urban areas, ranging from $38.7\%$ to $94.7\%$. Additionally, the voting trend is represented by historical voting patterns in presidential elections, categorizing each state as consistently voting for the Democratic or Republican candidate or being a Swing state.
We employ a Poisson regression model including one treatment-coded categorical predictor: the \emph{voting trend}. This predictor has three levels: Democrats, Republicans, and Swing, with Democrats serving as the reference category. Furthermore, the model incorporates one continuous predictor: the \emph{demographic trend}, measured as a percentage. The Poisson regression model is represented as follows:
\begin{align}
    \begin{split}
        y_i &\sim \textrm{Poisson}(\theta_i)\\
        \log(\theta_i) &= \beta_0 + \beta_1 x_1 + \beta_2 x_2 + \beta_3 x_3\\
        \beta_k &\sim \textrm{Normal}(\mu_k, \sigma_k) \quad \textrm{ for } k=0,\ldots,3.
    \end{split}
\end{align}
Here, $y_i$ is the number of counts for observation $i = 1, \ldots, N$. The counts follow a Poisson distribution with rate $\theta_i$. The rate parameter is predicted by a linear combination of two predictors: the continuous predictor $x_1$ with slope $\beta_1$ and a three-level factor represented by the coefficients $\beta_2$ and $\beta_3$ for both contrasts $x_2$ and $x_3$. The logged average count $y$ is denoted by $\beta_0$. All regression coefficients are assumed to have normal prior distributions with mean $\mu_k$ and standard deviation $\sigma_k$ for $k=0,\ldots,3$. The log-link function maps $\theta_i$ to the scale of the linear predictor. The main goal is to learn the hyperparameters $\lambda=(\mu_k, \sigma_k)$ based on expert knowledge.

\paragraph{Elicitation procedure}
The analyst queries the expert regarding two target quantities: the predictive distribution of the group means for states categorized as Democrats, Republicans, and Swing, and the expected number of LGBTQ+ anti-discrimination laws for selected US states $x_i$ with $i=1, 11, 15, 17, 27, 33$. Quantile-based elicitation is used for the distribution of group means and histogram elicitation for the observations per US state. Furthermore, the expert is queried about the expected maximum number of LGBTQ+ anti-discrimination laws in one US state. This information truncates the upper bound, $t^u$, of the Poisson distribution and is needed for applying the Softmax-Gumbel Trick which allows for computing gradients for discrete random variables (see Section~\ref{subsec:SGD} for details). For the current example, we assume $t^u=70$. The ideal expert is defined by the following true hyperparameters $\lambda^*=(\mu_0= 2.91, \sigma_0= 0.07, \mu_1= 0.23, \sigma_1= 0.05, \mu_2= -1.51, \sigma_2= 0.135, \mu_3= -0.61, \sigma_3= 0.105)$. The specification of algorithm parameters for the optimization procedure as well as a figure summarizing the convergence diagnostics can be found in Appendix \ref{app:diagnostics-pois}.

\begin{figure}[ht]
    \centering
    \includegraphics{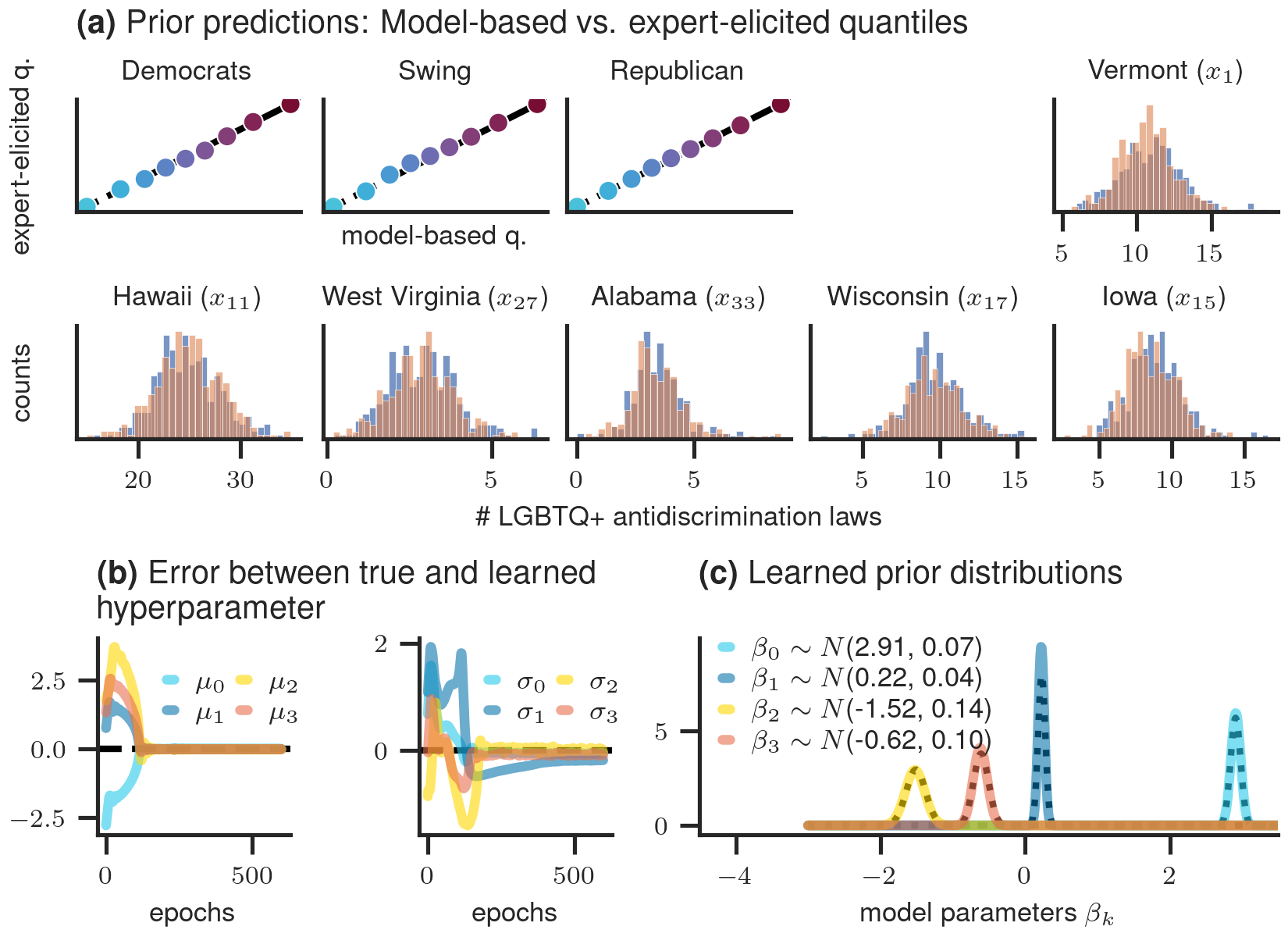}
    \caption{\emph{Results of Case Study 3}: \textbf{(a)} comparison between model-based and expert-elicited statistics. First three panels depict quantile-based elicitation for the group means, while the remaining upper panels show histogram elicitation for each state $x_i$. The model-based histograms are depicted in blue and the expert-elicited in red. \textbf{(b)} learning of hyperparameters across epochs, showcasing the difference between the true and learned values; \textbf{(a)} true (dotted line) and learned (solid line) prior distributions of the model parameters.}
    \label{fig:pois-sum-res}
\end{figure}

\paragraph{Results}
The learned hyperparameters' results are presented in Figure~\ref{fig:pois-sum-res}. In the upper panels, a comparison between model-based and expert-elicited statistics is presented. The level of agreement between model predictions and expert expectations is high for both elicitation formats: quantile-based elicitation for the voting groups in the first three panels and histogram elicitation for single states in the remaining upper panels. 
The model-based histograms are depicted in blue and the expert-elicited histograms in red. 
The lower left panels demonstrate that the error between learned and true hyperparameter values converges towards zero over time. Finally, the learned prior distributions are depicted in the lower right panel, with solid lines representing the learned and dotted lines the true priors.

An analysis of the learning algorithm's performance across various threshold values revealed that it functions effectively under different settings and maintains robustness to a large extent against both under- and overestimation of the actual threshold (refer to online supplement for details \url{https://osf.io/rxgv2/}). 

\subsection{Case Study 4: Hierarchical Models}
\label{subsec: case-study4}
\paragraph{Setup}
In this concluding case study, we investigate the performance of our elicitation method when applied to hierarchical models. This specific model class poses a distinct challenge for analysts and domain experts alike due to the inherent complexity of the model and the non-intuitive nature of varying effects (i.e., varying intercepts and slopes). Consequently, setting expectations for these model parameters from the perspective of domain experts becomes a difficult task. Our method allows for learning prior distributions within a hierarchical framework, while relying on expert knowledge that is articulated in terms of interpretable target quantities.

The accompanying example in this case study draws inspiration from the \emph{sleepstudy} dataset \supercite{belenky2003patterns} that comes along with the R-package \emph{lme4} \supercite{bates2015lme4}. This dataset contains information about the average reaction time (RT) in milliseconds for $N$ individuals who undergo sleep deprivation for nine consecutive nights (with less than three hours of sleep per night). In order to construct a model for this data, we consider a hierarchical model with days serving as a continuous predictor denoted as $x$,
\begin{align}\label{eq:mlm}
    \begin{split}
    y_{ij} &= \textrm{Normal}(\theta_{ij}, s)\\
    \theta_{ij} &= \beta_0 + u_{0,j} + (\beta_1 + u_{1,j}) x_{ij} \\
    ( u_{0,j}, u_{1,j} ) &\sim \textrm{MvNormal}\left(\textbf{0}, \Sigma_u\right) \\
    \Sigma_u &= \begin{pmatrix}\tau_0^2 & \rho_{01} \tau_0 \tau_1 \\ \rho_{01} \tau_0 \tau_1 & \tau_1^2\end{pmatrix} \\
    \beta_k &\sim \textrm{Normal}(\mu_k, \sigma_k) \qquad\textrm{ for } k=0,1\\
    \tau_k &\sim \textrm{TruncatedNormal}(0, \omega_k) \qquad\textrm{for } k=0,1\\
    \rho_{01} &\sim \textrm{LKJ}(\alpha_\textrm{LKJ}) \\
    s &\sim \textrm{Exponential}(\nu).
    \end{split}
\end{align}
Here $y_{ij}$ represents the average RT for the $j^{th}$ participant at the $i^{th}$ day with $j=1,\ldots, 300$ and $i=0,\ldots,9$. 
The RT data is assumed to follow a normal distribution with local mean $\theta_{ij}$ and within-person standard deviation $s$. Here, $\theta_{ij}$ is predicted by a linear combination of the continuous predictor $x$ with overall slope $\beta_1$ and intercept $\beta_0$. 
Given the potential variation in both baseline and change in RT across participants, the model incorporates varying (i.e., ``random'') intercepts $u_{0,j}$ and slopes $u_{1,j}$. These varying effects follow a multivariate normal distribution, centered at a mean vector of zero and with a covariance matrix $\Sigma_u$. This encodes the variability ($\tau_0, \tau_1$) and the correlation ($\rho_{01}$) between $u_{0,j}$ and $u_{1,j}$. 
For the resulting set of model parameters, the following prior distributions are assumed: A normal distribution for the overall (i.e., ``fixed'') effects $\beta_k$ ($k = 0, 1$) with mean $\mu_k$ and standard deviation $\sigma_k$. A truncated normal distribution centered at zero with a standard deviation of $\omega_k$, is employed for the person-specific variation $\tau_k$, which is constrained to be positive. The correlation parameter $\rho_{01}$ follows a Lewandowski-Kurowicka-Joe \supercite[LKJ; ][]{lewandowski2009generating} distribution with scale parameter $\alpha_\textrm{LKJ}$. In the subsequent context, we set  $\alpha_\textrm{LKJ}$ to 1. Additionally, an Exponential prior distribution with rate $\nu$ is used for the within-person (error) standard deviation $s$. The goal is to learn seven hyperparameters $\lambda=(\mu_k,\sigma_k, \omega_k, \nu)$ based on expert knowledge.

\paragraph{Elicitation procedure \& Optimization}
The analyst queries the expert regarding the following target quantities: the expected average RT for specific days $x_i$, where $i = 0,2,5,6,9$, the within-person standard deviation $s$ (elicitation in the parameter space), and the expected distribution of $R^2$ for the initial and final day ($i = 0,9$). We assume that the analyst employs quantile-based elicitation for the expected average RT per chosen day $x_i$. For the within-person standard deviation $s$, a moment-based elicitation approach is used, asking the expert regarding both the expected mean and standard deviation of $s$. Additionally, histogram-elicitation is utilized to query the expected $R^2$ distribution for each day. 
The ideal expert is defined by the following true hyperparameters: $\lambda^*=(\mu_0=250.40, \mu_1=30.26, \sigma_0=7.27, \sigma_1=4.82, \omega_0=33.00, \omega_1=23.00, \nu=0.04)$. Please refer to Appendix \ref{app:diagnostics-mlm1} for detailed information about the algorithm parameters of the optimization procedure together with a figure summarizing the convergence diagnostics indicating successful convergence.

\begin{figure*}[hpt!]
    \centering
    \includegraphics{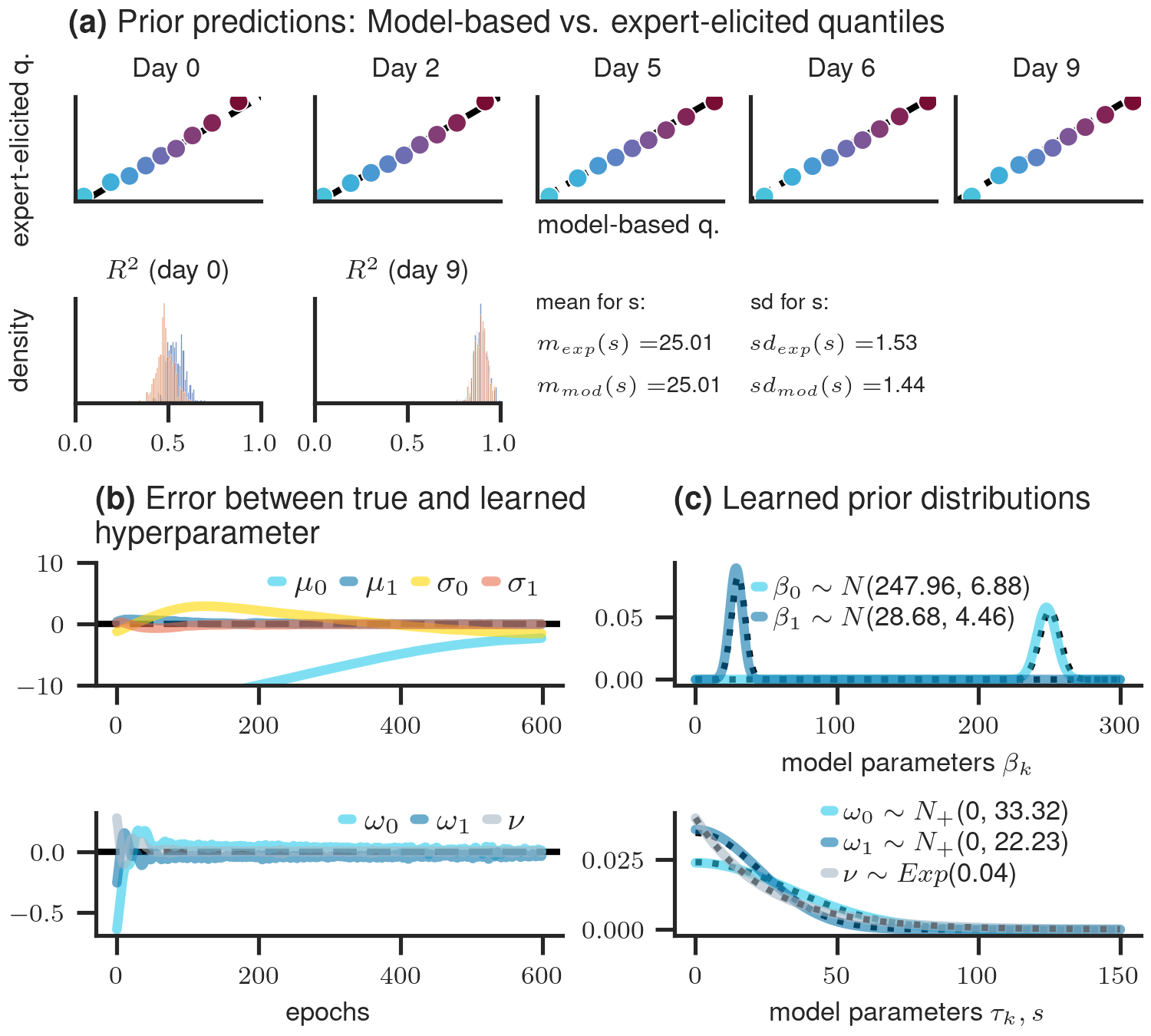}
    \caption{\emph{Results of Case Study 4 - Hierarchical normal model}: \textbf{(a)} comparison between model-based and expert-elicited statistics. First row depicts quantile-based elicitation for each day $x_i$. Second row shows histogram-based elicitation for $R^2$ (red expert and blue model-implied) and moment-based elicitation for model parameter $s$ ($m_\text{exp}, sd_\text{exp}$ stands for expert-elicited mean and standard deviation, respectively). \textbf{(b)} learning of hyperparameters across epochs, showcasing the difference between the true and learned values; \textbf{(c)} true (dotted line) and learned (solid line) prior distributions of each model parameter.}
    \label{fig:mlm-norm-sum-res}
\end{figure*}

\paragraph{Results} 
Figure~\ref{fig:mlm-norm-sum-res} presents the results derived from the optimization process. The upper two rows depict the congruence between simulation-based and expert-elicited statistics, effectively highlighting successful learning. The first row illustrates the alignment between expert-derived quantiles for the five chosen days $x_i$ and the corresponding simulated quantiles generated by the final trained model. The first two plots in the lower row show the distributions of $R^2$ as indicated by the expert and predicted by the model for day 0 and 9. The model-based histograms are depicted in blue and the expert-elicited in red. Finally, the queried moments (i.e., mean and standard deviation) for the model parameter $s$ are depicted as remaining information in the second row. 

The learned prior distributions for each model parameter are depicted in the lower, right column of Figure~\ref{fig:mlm-norm-sum-res}. The high overlap between true (dashed lines) and learned (solid lines) prior distributions indicates an additional instance of successful learning. This is further supported by the assessment of the error between true and learned hyperparameters in the lower left column, revealing a progressive convergence towards zero across epochs.

\section{Discussion}
\label{sec:discussion-conclusion}

When developing Bayesian models, analysts face the challenge of specifying appropriate prior distributions for each model parameter, involving both the choice of the distributional family as well as the corresponding hyperparameter values. 
We proposed an elicitation method that assists analysts in identifying the hyperparameter values of given prior distribution families based on expert knowledge. Our method accommodates various types of expert knowledge, including information about model parameters, observable data patterns, and other relevant statistics (e.g., $R^2$). Additionally, it can handle common expert data formats resulting from histogram, moment-, or quantile-based elicitation. 
Our method is agnostic to the specific probabilistic model and offers a modular design, providing analysts with the flexibility to modify or replace specific building blocks, such as the discrepancy measure or the loss weighting method.
In our case studies, we demonstrated the excellent performance of our method for various modeling tasks, combining different kinds of expert knowledge in a consistent and flexible manner. Despite these highly promising results, some relevant limitations remain, which are discussed below together with ideas for future research.

Our method employs gradient-based optimization to learn hyperparameter values. This choice offers the advantage of requiring only the ability to sample from the generative model, simplifying the learning process. However, it comes with the prerequisite that all operations and functions in the computational graph must be differentiable or admit a reparameterization whose gradients can be approximated with sufficient accuracy. Consequently, for discrete random variables, specific techniques, such as the Softmax-Gumbel trick that we used in our case studies, are necessary to obtain approximate gradients.
Alternatively, one could opt for optimization methods that entirely forego gradient computations. For instance, Manderson \& Goudie\supercite{manderson2023translating} propose a two-stage optimization process based on Bayesian optimization \supercite{frazier2018tutorial}, which avoids the need for gradients altogether. Nevertheless, this choice has its own limitations, notably in terms of scalability, as Bayesian optimization does not perform optimally in higher-dimensional spaces \supercite{eriksson2021high}.
Given the ongoing active research into the development of optimization techniques that can scale effectively in higher dimensions and handle discrete random variables, we acknowledge the potential for further advancement and refinement of our proposed method. 

Having a suitable optimization method is fundamental for learning hyperparameters based on expert knowledge. However, there are cases where hyperparameters cannot be uniquely determined from available expert data, leading to different learned hyperparameters upon multiple replications of the learning process. This situation raises the question of how to choose between prior distributions that represent the elicited expert knowledge equally well. Initial approaches, such as incorporating a regularization term in the loss function to favor priors with higher entropy, have been proposed to address this challenge Manderson \& Goudie\supercite{manderson2023translating}. Another avenue to achieve model identification involves the model architecture. For instance, statistical models that adopt joint priors for their parameters and thus keep the number of hyperparameters low, are expected to exhibit better model identification compared to models that assume independent priors for the parameters \supercite{aguilar2023intuitive}. Nevertheless, further research is needed to develop techniques that can efficiently handle unidentified models. Recently, Sameni\supercite{sameni2023beyond} emphasized the importance of establishing quantitative metrics for assessing model identification. We fully support this notion, as we also recognize the necessity for such metrics in our proposed method. Considering it as a future task, these metrics would provide valuable guidelines for analysts in making informed decisions regarding the selection among expert data collection strategies and model architectures. 

Finally, we need to discuss an important aspect shared by all gradient-based optimization methods, including our own: the objective of finding an optimal \emph{point estimate} for the hyperparameters $\lambda$. By adopting this approach, any uncertainties surrounding the value of $\lambda$ are neglected, despite the potential introduction of uncertainty during the prior elicitation process. For instance, an expert may possess uncertainties about the true value of a queried quantity or face uncertainty while quantifying implicit knowledge. The adoption of fixed point estimates for hyperparameters fails to account for these uncertainties and may result in an overly confident representation of the prior knowledge. 

To address this limitation, it would be advantageous to adopt a probabilistic approach that explicitly accounts for uncertainty in the hyperparameters. Some initial attempts to tackle this issue exist. For instance, Hartmann et al.\supercite{hartmann2020flexible} proposed modeling the uncertainty in quantifying expert knowledge using a Dirichlet distribution, where the precision parameter controls the variance. However, this approach is problem-specific, lacking generalizability to arbitrary models. An alternative probabilistic perspective to address this concern was suggested by Mikkola et al.\supercite{mikkola2023prior}, who advocate for a \emph{Bayesian treatment of the expert}. The central idea is to incorporate a user model for the expert and assume that the analyst holds a prior belief about the expert's knowledge, which is updated using Bayes' rule after each query. Given the flexibility of our method, it can readily accommodate this concept, offering a promising avenue for future development and next steps.

\newpage

\singlespacing
\printbibliography

\input{funding.tex}

\input{contributions.tex}
\input{data_availability}
\input{competing_interest}

\newpage
\thispagestyle{empty}
\listoffigures
\listoftables
\newpage
\pagenumbering{arabic}
      


\newpage
\section*{Appendix}
\input{appendix}


\end{document}

%% file: coverpage.tex

\begin{titlepage}

  \newcommand{\HRule}{\rule{\linewidth}{0.5mm}} 

  \center 

  \textsc{\Large Manuscript}\\[0.5cm] 
  \textsc{\large  }\\[0.5cm] 

  \vspace{1.5 cm}
  \HRule \\[0.4cm]
  { \huge \bfseries Simulation-Based Prior Knowledge Elicitation for Parametric Bayesian Models}\\[0.4cm] 
  \HRule \\[1.5cm]
 
  {\em\Large\textbf Authors:}\\
  \vspace{.5 cm}
  Florence \textsc{Bockting}*\\
  Department of Statistics\\
  TU Dortmund University, Germany\\
  \vspace{.5 cm}
  Stefan T. \textsc{Radev}\\
  Cognitive Science Department\\
  Rensselaer Polytechnic Institute, NY, USA\\
  \vspace{.5 cm}
  Paul-Christian \textsc{Bürkner}\\
  Department of Statistics\\
  TU Dortmund University, Germany\\

  \vspace{1.5 cm}
  {\large Last update: \today}\\[2.5cm]
  *\emph{corresponding author}: Florence Bockting, florence.bockting@tu-dortmund.de 

\vfill 
\end{titlepage}


\newpage

%% file: title_noauthors.tex

  \title{\vspace{-15mm}\fontsize{21pt}{10pt}\selectfont\textbf{Simulation-Based Prior Knowledge Elicitation for Parametric Bayesian Models}}

	\date{Submitted: \usvardate\today}

\makeatletter
\renewcommand{\@maketitle}{
\newpage
 \null
 \vskip 2em%
 \begin{center}%
  {\LARGE \@title \par}%
 \end{center}%
 \par} \makeatother

%% file: method_section.tex
\section{Methods}\label{sec:methods}
We propose a new elicitation method for translating knowledge from a domain expert into an appropriate parametric prior distribution. Building on recent contributions\supercite{hartmann2020flexible,da2019prior,manderson2023translating} we developed a model-agnostic method in which the search for appropriate prior distributions is formulated as an optimization problem. Thus, the objective is to determine the optimal hyperparameters that minimize the discrepancy between model-implied and expert-elicited statistics. Our elicitation method supports expert feedback in both the space of parameters and observable quantities (i.e., a \emph{hybrid} approach) 
and minimizes human effort. The key ideas underlying our method are outlined as follows:
\begin{enumerate}
    \item The analyst defines a generative model comprising a likelihood function $p(y\mid \theta)$ and a parametric prior distribution $p(\theta \mid \lambda)$ for the model parameters, where $\lambda$ represents the prior hyperparameters to be inferred from expert knowledge.
    \item The analyst selects a set of target quantities, which may involve queries related to observable quantities (data), model parameters, or anything else in between.
    \item The domain expert is queried using a specific elicitation technique for each target quantity (\textit{expert-elicited statistics}).
    \item From the generative model implied by likelihood, prior, and a given value of $\lambda$, parameters and (prior) predictive data are simulated, and the predefined set of target quantities is computed based on the simulations (\textit{model-implied quantities}).
    \item The discrepancy between the model-implied and the expert-elicited statistics is evaluated with a discrepancy measure (loss function). 
    \item Stochastic gradient descent is employed to update the hyperparameters $\lambda$ so as to minimize the loss function.
    \item Steps 4 to 6 are repeated iteratively until an optimal set of hyperparameters $\lambda$ is found that minimizes the discrepancy between the model-implied and the expert-elicited statistics.
\end{enumerate}
In the upcoming sections, we will delve into the details of the outlined approach. 
Following the terminology introduced in Mikkola et al.\supercite{mikkola2023prior}, our prior elicitation procedure involves two key individuals: the \emph{analyst} and the \emph{domain expert}. The analyst has a dual role, acting both as a statistician responsible for formulating the generative model and selecting the target quantities to be queried from the expert, as well as a facilitator who extracts the requested information from the domain expert. On the other hand, the domain expert possesses valuable knowledge pertaining to the uncertain quantities of interest that the analyst aims to extract. 
To provide a visual representation of all steps involved in our proposed elicitation method, Figure \ref{fig:concept} presents a graphical overview. In addition, readers can find a symbol glossary in Appendix~\ref{app:overview-param} for a quick reference. An illustrative example that details each step of the workflow using specific values can be found in our online supplement \url{https://osf.io/rxgv2}.
\begin{figure}[hpt!]
    \centering
    \includegraphics[width=\textwidth]{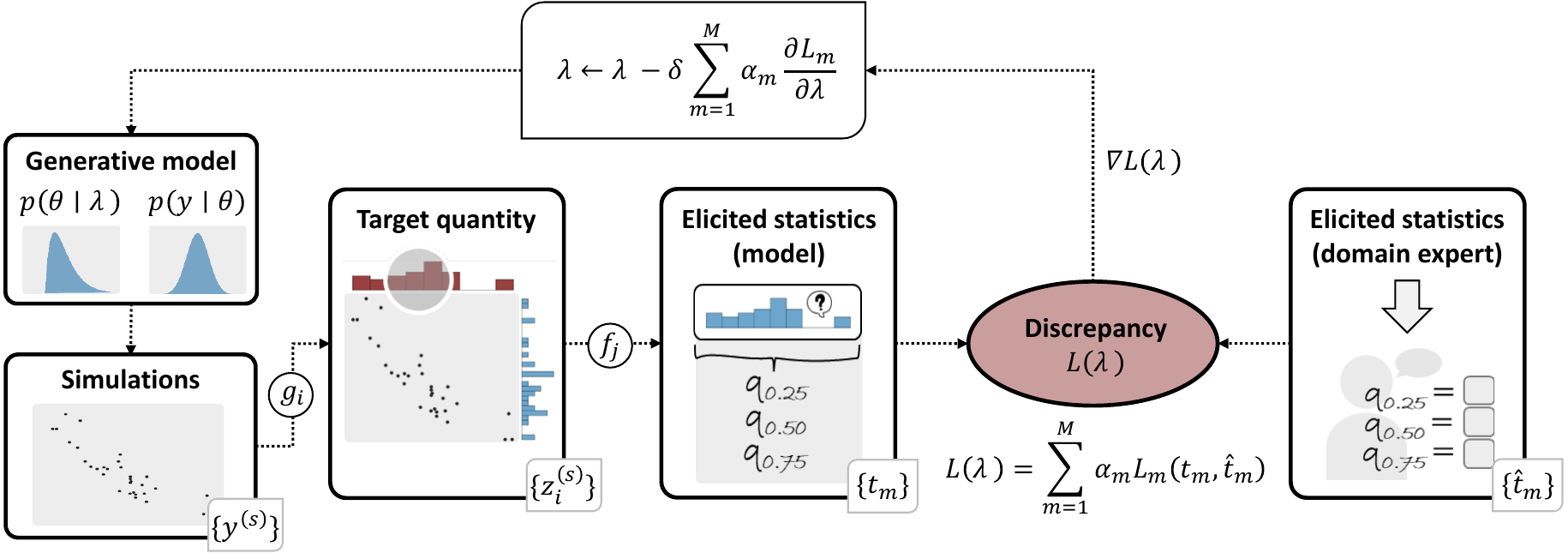}
    \caption{\emph{Graphical illustration of our simulation-based elicitation method.} Step 1 involves employing elicitation techniques to extract target quantities from the domain expert. Subsequently, the objective is to minimize the discrepancy between model-implied and expert-elicited statistics by optimizing the hyperparameters $\lambda$. The optimization process iteratively simulates data using the current hyperparameters $\lambda$, computes model-implied elicited statistics, compares them with the expert-elicited statistics using a loss function ($L_m$), and updates $\lambda$ to improve agreement between model-implied and expert-elicited statistics. Here, $\alpha_m$ is the weight of the $m^{th}$ loss component and $\delta$ is the step size.}
    \label{fig:concept}
\end{figure}

\subsection{Elicited Statistics From the Expert}\label{subsec:expert-quant}
We assume that the analyst queries the domain expert regarding a predetermined set of $I$ target quantities, represented as $\{z_i\} := \{z_i\}_{i=1}^I$. The set $\{z_i\}$ is selected by the analyst, whereby the choice of target quantities is influenced by the requirements of the statistical model \supercite{da2019prior,stefan2022practical}. Additionally, the selected target quantities should align with the expert's knowledge, encompassing both potential knowledge about model parameters and/or observable data patterns \supercite{mikkola2023prior}.
Once this set is defined, the expert is queried regarding each individual target quantity $z_i$, assuming that the expert possesses an implicit representation, denoted as $\hat{z}_i$, which can be accessed using expert elicitation techniques \supercite{manderson2023translating, kadane1998experiences}.

The utilization of \emph{elicitation techniques} for querying the expert is essential for effectively capturing the probabilistic nature of the expert's judgment \supercite{o2006uncertain,hartmann2020flexible}. While numerous elicitation techniques have been proposed in the literature \supercite{stefan2022practical}, it can be argued that these techniques essentially represent different facets of the following three general method families: moment-based elicitation (e.g., mean and standard deviation), quantile-based elicitation (e.g., median, lower quartile, and upper quartile), and histogram elicitation (e.g., constructing a histogram by sampling from the distribution of $z_i$). Each target quantity $z_i$ can be elicited through a distinct elicitation technique $f_j$.
As emphasized by Falconer et al.\supercite{falconer2022methods} and Albert et al.\supercite{albert2012combining}, the selection of an appropriate elicitation technique depends on the specific context and must align with the expert's preferences. Hence, it is desirable to possess a methodology that accommodates different encoding formats. Within our notation, we represent the $i^{th}$ target quantity elicited from the expert through the $j^{th}$ elicitation technique as $\hat{t}_m = \hat{t}_{ij}$ and refer to it as \emph{elicited statistics}:
\begin{align}
\hat{t}_m = f_j(\hat{z}_i).
\end{align}
In this context, the index $m = 1, \ldots, M$ indicates the number of elicited statistics resulting from specific target-quantity $\times$ elicitation-technique combinations, as selected by the analyst. We use $m$ instead of $ij$ as subscript in order to emphasize that the set of elicited statistics consists of a \emph{selection} rather than all possible combinations $ij$. 

\subsection{Model-Based Quantities}\label{subsec:model-quant}
Considering the set of elicited statistics queried from the expert $\{\hat{t}_m\}$, it is possible to assess the extent to which a generative model, as specified by the analyst, aligns with the expert's expectations. A Bayesian model comprises a likelihood $p(y \mid \theta)$ as well as parametric prior distributions $p(\theta \mid \lambda)$ for the model parameters $\theta$. Here, $\lambda$ represents the prior hyperparameters to be inferred by our method and $y$ denotes a vector of observations.
The prior distributions incorporate prior information into the model. Thus, the degree to which the model captures the expert's expectations relies on the specific values assigned to $\lambda$. 
Consequently, the objective is to identify an appropriate specification of $\lambda$ that minimizes the discrepancy between the set of elicited statistics from the expert $\{\hat{t}_m\}$ and a corresponding set of elicited statistics derived from the model, $\{t_m\}$, to be discussed next. 

First, we need to derive the set of model-implied target quantities $\{z_i\}$. As a target quantity can represent an observable, a parameter, or anything else in between, we define it in the most general form as a function of the model parameters $\theta$, denoted as $z = g(\theta)$, where the function $g$ can take on various forms and be of deterministic or stochastic nature.
In its simplest form, the target quantity directly corresponds to a parameter of interest in the data-generating model ($z_i=g(\theta) = \theta_i$; i.e., $g$ would be a simple projection). 
Alternatively, $g$ can be aligned with the generative model of the data, resulting in the target quantity being equivalent to the observations ($z_i = g(\theta) = y$).

Moreover, the function $g$ can take on more complex forms. Suppose the domain expert provides prior knowledge about the coefficient of determination $R^2$. The $R^2$ measure expresses the proportion of variance explained by the model in relation to the total variance of the response $y$ and is commonly used to measure model fit in regression models \supercite{gelman2019r}. To obtain the corresponding model-implied $R^2$, we first generate observations $y$ using the specified generative model and then compute the $R^2$ value from the observations. 
Given the set of model-implied target quantities, we get the respective model-implied \emph{elicited statistics}, denoted by $\{t_m\}$, by applying the elicitation technique $f_j$ to the target quantity $z_i$:
\begin{align}
t_m = f_j(z_i).
\end{align}
A challenge with this approach is that the distribution of $\{t_m\}$ may not be analytical or have a straightforward computational solution. For instance, consider the case where the target quantity is equivalent to the observations, $z_i = y$. In this case, the distribution of the predicted observations $y$ gives rise to an integral equation known as the prior predictive distribution (PPD), denoted by $p(y \mid \lambda)$ and defined by averaging out the prior from the generative model:
\begin{equation}\label{eq:PPD-def}
p(y \mid \lambda) = \int_\Theta p(y \mid \theta) p(\theta \mid \lambda) d\theta.
\end{equation}
Obtaining a closed-form expression for this integral is only feasible in certain special cases, such as when dealing with conjugate priors. This challenge extends to all situations where the target quantity is a function of the observations $y$. Additionally, as previously mentioned, the \emph{elicited} statistic is defined as a function of the target quantity using an elicitation technique $f_j$ which involves the computation of moments or quantiles, among other possibilities. Consequently, the resulting form of the elicited statistic may once again pose analytic challenges.

Our primary objective is to ensure the broad applicability of our elicitation method to a wide range of models, including those with intractable likelihood functions \supercite{cranmer2020frontier}. To achieve this, we adopt a simulation-based approach that relies solely on the ability to generate samples from the relevant quantities. Bayesian models, by their very formulation, can simulate data from their prior and likelihood distributions, thereby enabling us to generate samples from the Bayesian probabilistic model \supercite{aushev2023online,gelman2013bayesian}. 
For example, in the case where $z_i = y$, the simulation-based procedure involves two steps: Firstly, we sample the model parameters from the prior distribution conditioned on hyperparameters $\lambda$:  $\theta^{(s)} \sim p(\theta \mid \lambda)$. Subsequently, we generate data by sampling from the likelihood distribution, resulting in $y^{(s)} \sim p(y \mid \theta^{(s)})$. The superscript ${(s)}$ is used to denote the $s^{th}$ sample of the corresponding simulated quantity. By repeating these steps, we can generate a collection of $S$ simulations $\{y^{(s)}\} := \{y^{(s)}\}_{s=1}^S$, where each element corresponds to a data point drawn from the PPD
\begin{equation}\label{eq:PPD-samples}
    y^{(s)} \sim p(y\mid \lambda).
\end{equation}
This approach is equivalent to Monte Carlo approximation of the integral in the PPD (Eq.~\ref{eq:PPD-def}), since it yields random evaluation points $y^{(s)}$ from the ``marginal'' distribution $p(y\mid \lambda)$.

\subsection{Multi-Objective Optimization Problem}\label{subsec:moop}
Once the elicited statistics $\{\hat{t}_m\}$ from the expert and a procedure to compute the corresponding model-implied elicited statistics $\{t_m\}$ are chosen, the focus can be shifted towards the main objective: Determine the hyperparameters $\lambda$ that minimize some discrepancy measure (loss function) $L(\lambda)$ between the expert elicited $\{\hat{t}_m\}$ and the model-implied statistics $\{t_m\}$ = $\{t_m(\lambda)\}$. Since the evaluation of the discrepancy extends to all elicited statistics $\{t_m\}$, $L(\lambda)$ has to be formulated as a multi-objective loss function. This loss function encompasses a linear combination of discrepancy measures $L_m$, with corresponding weights $\alpha_m$ (see Section~\ref{subsec:DWA}). In the following, we will also use the term \emph{loss components} to refer to the individual components in the weighted sum. The selection of the discrepancy measure $L_m$ is contingent upon the elicited statistic, therefore different choices may be appropriate depending on the specific quantity to be compared (see Section~\ref{subsec:MMD}). Independently of these specific choices, our main objective can be written as
\begin{equation}\label{eq:objective-single}
    \lambda^* = \arg \min_{\lambda} L(\lambda) = \arg \min_{\lambda} \sum_{m=1}^M\alpha_{m} L_m(t_m(\lambda), \hat{t}_m),
\end{equation}
where $\lambda^*$ denotes the optimal value of the hyperparameters $\lambda$ given the provided expert knowledge.

\subsection{Gradient-based Optimization}\label{subsec:SGD}
The optimization procedure for solving Equation (\ref{eq:objective-single}) follows an iterative approach. In each iteration, we sample from the generative model, compute the model-implied elicited statistics, and update the hyperparameters $\lambda$. This update relies on calculating the gradient of the discrepancy loss with respect to the hyperparameters $\lambda$ and adjusting them in the opposite direction of the gradient \supercite{goodfellow2016deep}. The procedure continues until a convergence criterion is met, usually when all elements of the gradient approach zero. We employ mini-batch stochastic gradient descent (SGD) with automatic differentiation, facilitated by the (explicit or implicit) reparameterization trick\supercite{kingma2014auto,figurnov2018implicit}. In our case, stochasticity in mini-batch SGD arises naturally as we simulate new model-implied quantities at each iteration step. 

The reparameterization trick involves splitting the representation of a random variable into stochastic and deterministic parts. By differentiating through the deterministic part, we can compute gradients with respect to $\lambda$ using automatic differentiation \supercite{sjolund2023tutorial}.
To leverage backpropagation, it is essential that all operations and functions in the computational graph are differentiable with respect to $\lambda$. This requirement extends to the loss function and all computational operations in the generative model \supercite{hartmann2020flexible, da2019prior}.

However, dealing with discrete random variables poses a challenge due to the non-differentiable nature of discrete probability distributions, making gradient descent through such variables difficult. One approach to overcome this challenge is to use continuous relaxation of discrete random variables, which enables the estimation of gradients and thus the use of gradient-based optimization methods for models that involve discrete random variables \supercite{tokui2016reparameterization,maddison2016concrete,jang2016categorical}.  
For instance, both Maddison et al.\supercite{maddison2016concrete} and Jang et al.\supercite{jang2016categorical} independently proposed the Gumbel-Softmax trick, which approximates a categorical distribution with a continuous distribution.
Let $x \sim \textrm{Categorical}(\boldsymbol\pi)$, where $\pi_i$ represents the probability of category $i$ among $n$ categories. The Gumbel-Max trick \supercite{gumbel1962statistical,maddison2014sampling} provides a simple and efficient way to draw samples from a categorical distribution by setting $x_i = \textrm{one-hot}(\arg \max_i [g_i + \log \pi_i])$, where $g_i$ denotes an iid sample from a Gumbel$(0,1)$ distribution  \supercite{jang2016categorical}.  However, the derivative of the $\arg \max$ is 0 everywhere except at the boundary of state changes, where it is undefined. To overcome this limitation, Jang et al.\supercite{jang2016categorical} and Maddison et al.\supercite{maddison2016concrete} independently proposed using the softmax function as a continuous, differentiable approximation to the $\arg \max$ operator, and generating $k-$dimensional vectors in a $(k-1)$-dimensional simplex $\Delta^{k-1}$ where
\begin{equation}\label{eq:Gumbel-Softmax}
    x_i = \frac{\exp\left\{(\log \pi_i + g_i)\tau^{-1}\right\}}{\sum_j \exp\left\{(\log \pi_j + g_j)\tau^{-1}\right\}}.
\end{equation}
Here, $\tau$ denotes the temperature hyperparameter, whereby small temperature values reduce the bias but increase the variance of gradients. In the case of $\tau\rightarrow 0$, the Gumbel-Softmax distribution is equivalent to the categorical distribution \supercite{jang2016categorical}. Higher temperature values lead to smoother samples, reducing the variance of gradients but introducing more bias. We observed robust results for $\tau = 1.0$ in our simulations and used it for all case studies. 

The Gumbel-Softmax trick is applicable to discrete distributions with a limited (finite) number of categories. Recently, Joo et al.\supercite{joo2021generalized} proposed an extension of the Gumbel-Softmax trick to arbitrary discrete distributions by introducing truncation for those distributions that lack upper and/or lower boundaries. This extension introduces an extra degree of freedom in the form of specifying the truncation level. A wider truncation range brings the approximate distribution closer to the original distribution but also increases the associated computational costs. For a detailed algorithm outlining the Gumbel-Softmax trick with truncation, please refer to Joo et al.\supercite{joo2021generalized}.
In our study, we applied the truncation technique for the Poisson distribution in Case Study 3 (Subsection \ref{subsec: case-study3}).

\subsection{Maximum Mean Discrepancy}\label{subsec:MMD}
A key aspect of the optimization problem, as expressed in Equation (\ref{eq:objective-single}), is the selection of an appropriate discrepancy measure, $L_m$. This measure depends on the characteristics of the elicited statistics $\{t_m\}$ and $\{\hat{t}_m\}$. Given that our method entails the generation of $\{t_m\}$ through repeated sampling from the generative model, a loss function is needed that can quantify the discrepancy between samples. The \emph{Maximum Mean Discrepancy}\supercite{gretton2008kernel,gretton_2017} (MMD) is a kernel-based method designed for comparing two probability distributions when only samples are available, making it suitable for our specific requirements. We utilize the MMD for all loss components in our applications. This decision is based on the robust simulation results and excellent performance reported in the Case Studies section. That said, our method does not strictly require the MMD, but allows analysts to choose a different discrepancy measure for each loss component, if desired.

Let $x= \{x_1, \ldots, x_n \}$ and $y= \{y_1, \ldots, y_m \}$ be independently and identically distributed draws from the distributions $p$ and $q$, respectively. 
The MMD is defined as the distance between the two distributions $p$ and $q$ and can be expressed as follows\supercite{gretton2008kernel}
\begin{align}\label{eq:MMD-def1}
    \textrm{MMD}[\mathcal{F},p,q] = \sup_{f \in \mathcal{F}}\left(\mathbb{E}_{x\sim p}[f(x)] - \mathbb{E}_{y\sim q}[f(y)] \right)
\end{align}
where $\mathcal{F}$ is the set containing all continuous functions. If this set is a unit ball in the universal \emph{reproducing kernel Hilbert space} (RKHS) $\mathcal{H}$ with associated reproducing kernel $k(\cdot,\cdot)$, the $\textrm{MMD}[\mathcal{H},p,q]$ is a strictly proper divergence, so it equals zero if and only if $p=q$. Intuitively, $\textrm{MMD}[\mathcal{H},p,q]$ is expected to be small if $p \approx q$, and large if the distributions are far apart \supercite{gretton_2017}. 
The (biased) empirical estimate of the squared-MMD is defined as
\begin{equation}
    \textrm{MMD}^2_b[\mathcal{H}, p,q] = \frac{1}{n^2}\sum_{i,j = 1}^n k(x_i, x_j) + \frac{1}{m^2}\sum_{i,j = 1}^m k(y_i,y_j) - \frac{2}{mn} \sum_{i,j=1}^{m,n} k(x_i,y_j)
\end{equation}
where $k(\cdot, \cdot)$ is a continuous and characteristic kernel function.
One popular choice for a characteristic kernel is the \emph{Gaussian} kernel \supercite{muandet2017kernel}, defined as $k(x,y) = \exp\{-||x-y||^2/ (2\sigma^2 )\}$ where $\sigma^2$ represents the bandwidth. To mitigate sensitivity to different choices of the bandwidth $\sigma^2$, a common approach is to use a mixture of $G$ Gaussian kernels that cover a range of bandwidths, given by $k(x,y) = \sum_{i=1}^G k_{\sigma_i}(x,y)$ where $\sigma_i$ denotes the bandwidth for the respective Gaussian kernel\supercite{li2015generative,schmitt2021detecting,feydy2020geometric}. 
In our simulations, we used the \emph{energy distance} kernel $k(x,y)= - ||x-y||$, as proposed by Feydy\supercite{feydy2020geometric} and Feydy et al.\supercite{feydy2019interpolating}, which does not require an extra hyperparameter for tuning. 
In our simulations, presented in the case studies section, we examined both kernels. However, due to the robustness of the results obtained with the energy kernel, along with its additional advantages of faster learning and the absence of an extra hyperparameter to be specified, we employed the energy kernel exclusively for all experiments presented in this paper.

\subsection{Dynamic Weight Averaging}\label{subsec:DWA}
\label{subsec:weights}
Formulating the multi-objective optimization problem as a weighted sum of loss functions necessitates an important consideration regarding the choice of weights $\alpha_m$ in Equation (\ref{eq:objective-single}). One possibility is for the user to customize the choice of $\alpha_m$, signifying the varying degrees of importance for each loss component in a particular application \supercite{deb2011multi}. However, another consideration refers to the \emph{task balancing problem}. When employing stochastic gradient descent to minimize the objective outlined in Equation~(\ref{eq:objective-single}), the hyperparameters $\lambda$ are updated according to the following rule
$\lambda \leftarrow \lambda - \delta \sum_{m=1}^M \alpha_m \frac{\partial L_m}{\partial \lambda},$ where $\delta$ is the step size (i.e., learning rate).
The equation suggests that the hyperparameter update may not yield optimal results if one loss component significantly outweighs the others \supercite{deb2011multi}. 

Consequently, a strategy is needed to dynamically modify the weights $\alpha_m$ to ensure effective learning of all loss components. The \emph{Dynamic Weight Averaging} (DWA) method proposed by Liu et al.\supercite{liu2019end} determines the weights based on the learning speed of each component, aiming to achieve a more balanced learning process. Specifically, the weight of a component exhibiting a slower learning speed is increased, while it is decreased for faster learning components \supercite{crawshaw2020multi}. 
We employ a slightly modified variant of DWA, where the weight $\alpha_m$ for loss component $m$ at current step $t_\textrm{curr}$ is set as
\begin{align}\label{eq:DWA}
    \alpha_m^{(t_\textrm{curr})} &= \frac{M \cdot \exp(\gamma_m^{t_\textrm{prev}}/a)}{\sum_{m'=1}^M \exp(\gamma_{m'}^{t_\textrm{prev}}/a)}\quad\textrm{ with }\quad 
    \gamma_m^{t_\textrm{prev}} =\frac{L_m^{t_\textrm{prev}}}{L_m^{t_\textrm{start}}}.
\end{align}
Here, $M$ is the total number of loss components $L_m$, $t_\textrm{prev}$ indexes the previous iteration step and $t_\textrm{start}$ refers to the initial iteration step, $\gamma_m$ calculates the relative rate of descent, and $a$ is a temperature parameter that controls the softness of the loss weighting in the softmax operator. Setting the temperature $a$ to a large value results in the weights $\alpha_m$ approaching unity. In our case studies (see Section \ref{sec:case-studies}), we obtained favorable outcomes when employing a value of around $a=1.6$, which we utilized consistently across all case studies.

The definition of DWA in Equation (\ref{eq:DWA}) includes a slight modification compared to its original formulation, specifically concerning the rate of change of individual loss components, $L_m$.
While Liu et al.\supercite{liu2019end} compare the loss ratio between two consecutive iteration steps, we have adopted the definition from the \emph{Loss-Balanced Task Weighting} method proposed by Liu et al.\supercite{liu2019loss}, where the loss ratio is calculated by comparing the current loss to the initial loss. This modification is motivated by the stochastic fluctuations of the loss between consecutive steps. However, the loss ratio between the current and initial loss can be a useful proxy for assessing the model's training progress for a given task \supercite{liu2019loss}. 
While the loss values of two consecutive SGD iterations may exhibit strong fluctuations, the overall trend should follow a decrease in the loss value as the algorithm converges. By comparing the current loss to the initial loss, we can then gain insights into how well a particular task has been learned. 
Tasks that are poorly learned would have ratios close to 1, indicating their larger contribution to the overall loss and gradient.

%% file: funding.tex
\section*{Acknowledgements:}
FB and PCB were supported by Deutsche Forschungsgemeinschaft (DFG, German Research Foundation) under Germany’s Excellence Strategy -- EXC-2075-390740016 (the Stuttgart Cluster of Excellence SimTech). STR war supported by the Deutsche Forschungsgemeinschaft (DFG, German Research Foundation) under Germany’s Excellence Strategy -- EXC-2181-390900948 (the Heidelberg Cluster of Excellence STRUCTURES).

%% file: contributions.tex
\section*{Contributions:}
FB and PCB drafted the manuscript and all authors contributed to writing the manuscript. All authors provided revisions to the manuscript, discussed the results and commented on the manuscript. PCB supervised the research.

%% file: data_availability.tex
\section*{Data \& Code availability:}
All code and data is openly available on OSF \url{https://osf.io/rxgv2}.

%% file: competing_interest.tex
\section*{Competing interests:}
The authors declare no competing interests.

%% file: appendix.tex
\pagenumbering{roman} 
\renewcommand{\thesubsection}{\Alph{subsection}}
\renewcommand\thesection{\Roman{section}}
\renewcommand\thefigure{\thesubsubsection.\arabic{figure}} 
\renewcommand\thetable{\thesubsubsection.\arabic{table}} 
\setcounter{figure}{0}
\setcounter{table}{0}
\setcounter{subsection}{0}
\subsection{Method}\label{app:overview-param}
\input{tables/table-1}

\subsection{Case Studies: Convergence diagnostics and Learning Results}
\subsubsection{Case Study 2: LM --- Normal linear regression model}\label{app:diagnostics-linreg}
To instantiate the optimization process the hyperparameter $\lambda$ are randomly initialized: $\mu_k \sim \text{Normal}(0,0.1)$, $\log \sigma_k \sim \text{Uniform}(-2, -4)$, and $\log \nu \sim \text{Uniform}(1, 2)$. To enhance learning behavior, we utilize a reparameterization trick for $s \sim \textrm{Exponential}(\nu)$, achieved by expressing a prior on the mean of $M$ replicated $s$ as $1/M \sum_n s_m \sim \textrm{Gamma}(M, M \cdot \nu)$. The algorithm parameters for the optimization procedure were set as follows: $B = 2^8$, $E=400$, $S=300$, and an exponential learning rate schedule that decays every $5$ steps with a base of $0.90$, an initial learning rate $\phi^0=0.01$, and $\phi^{\min}= 10^{-2}$. In total the learning algorithm need 20.76 minutes to finish. Figure with convergence diagnostics of the optimization results can be found in the Case Studies section in the main text.

\subsubsection{Case Study 2: GLMs --- Binomial model}\label{app:diagnostics-binom}
\begin{figure*}[ht]
    \centering
    \includegraphics{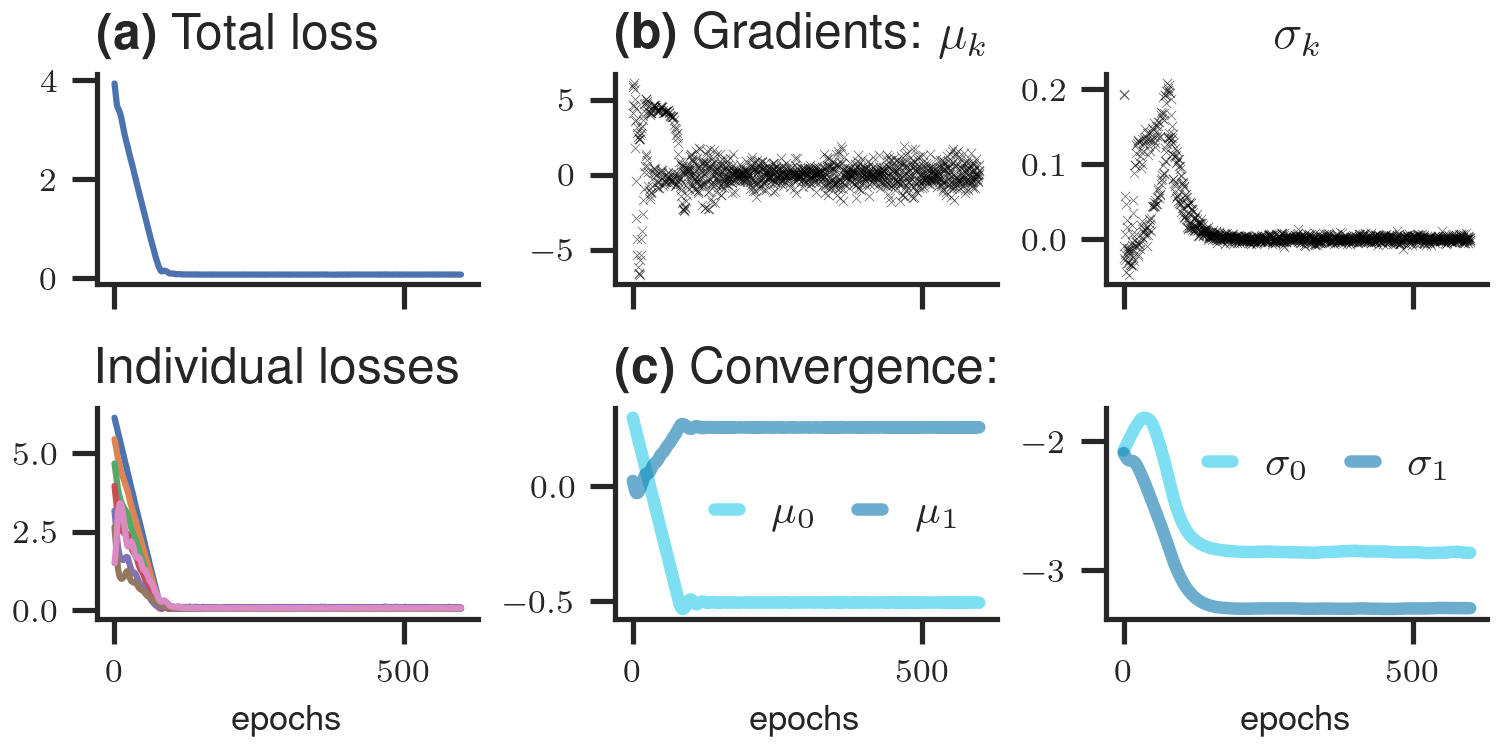}
    \caption{\emph{Convergence diagnostics Binomial model.} \textbf{(a)} loss value across epochs, demonstrating the desired decreasing trend of all loss values (i.e., total loss as well as individual loss components); \textbf{(b)} gradients of the hyperparameters $\lambda$ show decreasing trend towards zero; \textbf{(c)} learning of each hyperparameter across epochs, stabilizing in the long run at a specific value.}
    \label{fig:conv-diag-binom}
\end{figure*}

To setup the learning algorithm, the following algorithm parameters are used: $B=2^8$, $E=600$, $S=300$, and an exponential learning rate schedule that decays every $5$ steps with a base of $0.90$. The initial learning rate is set to $\phi^0=0.01$, and the minimum learning rate is $\phi^{\min}=10^{-5}$. To instantiate the optimization process the hyperparameter $\lambda$ are randomly initialized: $\mu_k \sim \text{Normal}(0,1)$ and $\log \sigma_k \sim \text{Uniform}(-2, -3)$. To improve gradient-based learning, we divide the continuous predictor by its standard deviation and use this scaled predictor in the optimization process. In total the learning algorithm need 23.28 minutes to finish. Figure \ref{fig:conv-diag-binom} depicts the convergence diagnostics of the optimization results.

\setcounter{figure}{0}
\setcounter{table}{0}
\subsubsection{Case Study 3: GLMs --- Poisson model}  \label{app:diagnostics-pois}
\begin{figure*}[ht]
    \centering
    \includegraphics{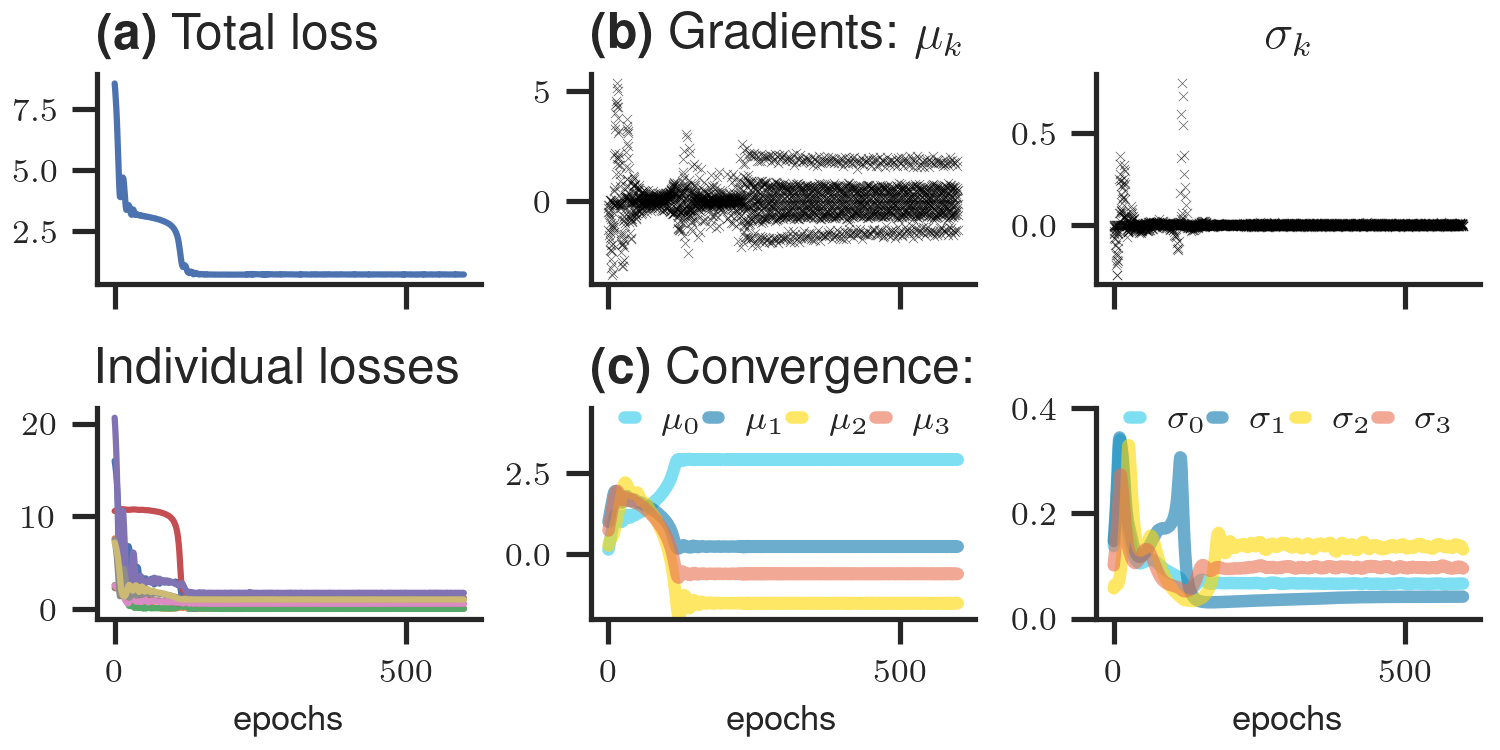}
    \caption{\emph{Convergence diagnostics for Poisson model.} \textbf{(a)} decreasing trend of the loss across epochs  (i.e., upper panel total loss and lower panel individual loss components); \textbf{(b)} gradients of the hyperparameters $\lambda$ across epochs; \textbf{(c)} updating of each hyperparameter across epochs.}
    \label{fig:conv-diag-pois}
\end{figure*}

For the learning algorithm, we standardize (z-transform) the continuous predictor to improve learning and we set the algorithm parameters as follows: $B=2^8, E=600$, $S=300$, and an exponential learning rate schedule that decays every $5$ steps with a base of $0.90$, an initial learning rate $\phi^0=0.1$, and $\phi^{\min}=10^{-3}$. To instantiate the optimization process the hyperparameter $\lambda$ are randomly initialized: $\mu_k \sim \text{Uniform}(0,1)$ and $\log \sigma_k \sim \text{Uniform}(-2, -3)$. In total the learning algorithm need 40.03 minutes to finish. The convergence diagnostics inspection shows successful convergence. Figure \ref{fig:conv-diag-pois} depicts the convergence diagnostics of the optimization results.
\newpage
\setcounter{figure}{0}
\setcounter{table}{0}

\subsubsection{Case Study 4: Hierarchical model --- Normal likelihood} \label{app:diagnostics-mlm1}
\begin{figure*}[ht]
    \centering
    \includegraphics[width=0.8\textwidth]{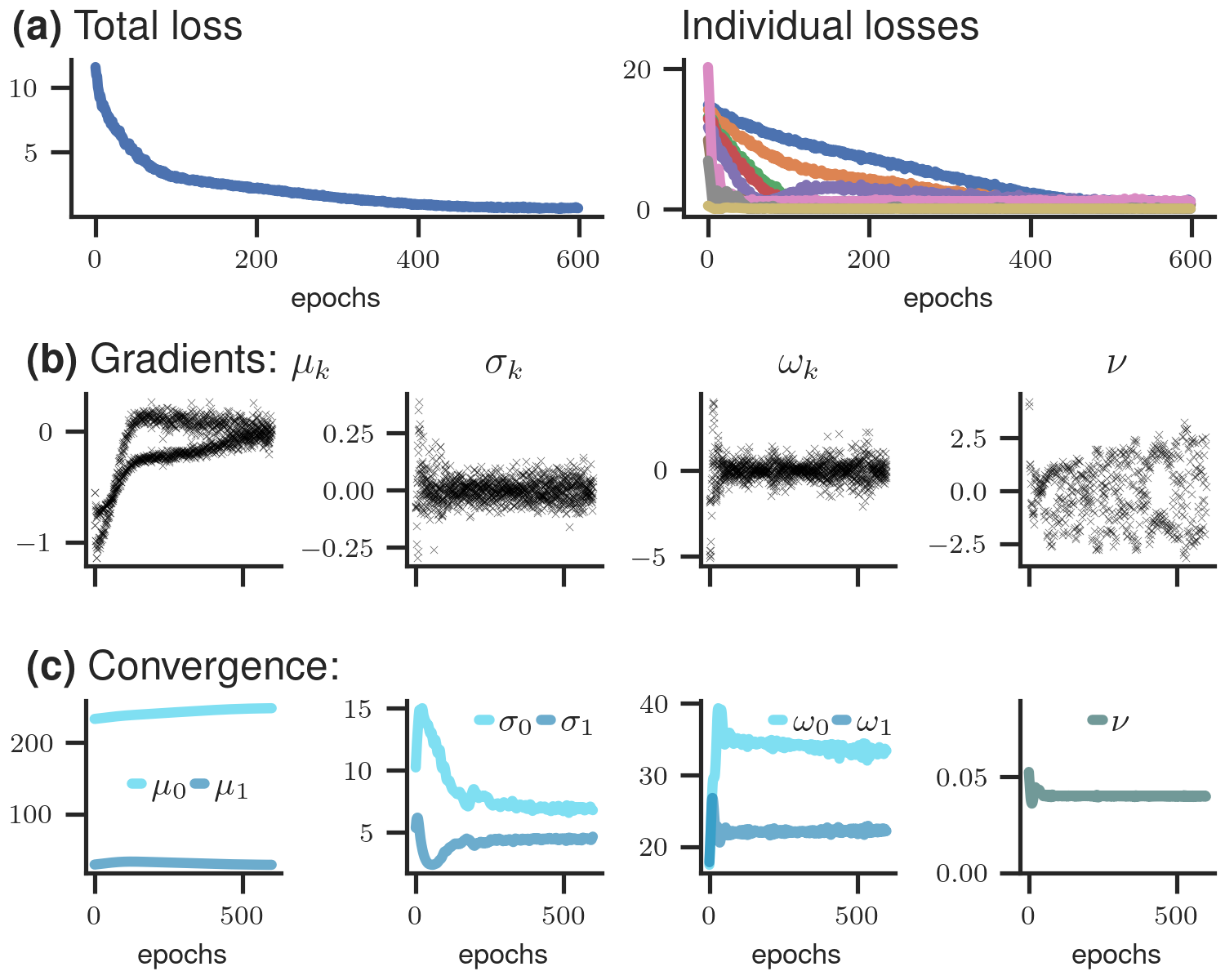}
    \caption{\emph{Convergence diagnostics for multilevel model with normal likelihood.} \textbf{(a)} loss value across epochs, demonstrating the desired decreasing trend of all loss values (i.e., total loss on the left as well as individual loss components on the right); \textbf{(b)} gradients of each learned hyperparameter $\lambda$; \textbf{(c)} learning of each hyperparameter across epochs.}
    \label{fig:conv-diag-mlmNorm}
\end{figure*}

For the learning algorithm, we divide the continuous predictor by its standard deviation and use this scaled predictor in the optimization process. We set the algorithm parameters as follows: $B=2^7, E=600$, $S=300$, and an exponential learning rate schedule that decays every $5$ steps with a base of $0.90$, an initial learning rate $\phi^0=0.05$, and $\phi^{\min}=10^{-2}$. 
To enhance learning behavior, we utilize a reparameterization trick for $s \sim \textrm{Exponential}(\nu)$, achieved by expressing a prior on the mean of $M$ replicated $s$ as $1/M \sum_n s_m \sim \textrm{Gamma}(M, M \cdot \nu)$.
To instantiate the optimization process the hyperparameter $\lambda$ are randomly initialized: $\mu_0 \sim \text{Normal}(240,5), \mu_1 \sim \text{Normal}(40,5), \log\sigma_0 \sim \text{Normal}(2, 0.5), \log\sigma_1 \sim \text{Normal}(1.5, 0.5), \log\omega_k \sim \text{Normal}(3, 0.5)$ and $\log \nu \sim \text{Uniform}(-2, -3)$. 
In total the learning algorithm need 53.08 minutes to finish.
The convergence diagnostics inspection shows successful convergence. Figure~\ref{fig:conv-diag-mlmNorm} depicts a summary of convergence diagnostics.

%% file: tables/table-1.tex
\begin{table}[hpt!]
    \subsubsection{Symbol Glossary}
    \caption{Overview of model and algorithm (hyper-)parameters}
    \centering
    \begin{tblr}{
    p{0.18\textwidth}p{0.1\textwidth}p{0.2\textwidth}p{0.3\textwidth}
    }
     \hline
     Workflow task & Notation & Label & Comment \\
     \hline
    General & seed & seed & seed = 2023 \\
    \hline[dashed]
    \SetCell[r=2]{l}Generative Model & $\lambda$ & model hyperparameter & hyperparameter of prior distributions\\
    & $\theta$ & model parameter & model parameter \\
    \hline[dashed]
    Dynamic Weight Averaging & $a$ & temperature & $a=1.6$ \\
    \hline[dashed]
    Gumbel-Softmax Trick & $\tau$ & temperature & $\tau = 1.0$ \\
    \hline[dashed]
    \SetCell[r=3]{l}Gradient Descent & $E$ & epochs / iterations & number of iterations used for learning the model hyperparameter \\
    & $B$ & batch size & number of simulations within one epoch; fix for all case studies\\
    & $S$ & prior-samples & number of samples from the prior distribution of each model-parameter within one batch \\
    \hline[dashed]
    \SetCell[r=4]{l}Adam Optimizer (with exponential decay) & $\textrm{lr}_0$ & initial learning rate & learning rate of the initial iterations \\
    & $\textrm{lr}_{\min}$ & minimum learning rate & lower bound for the decay \\
    & $\textrm{lr}_{\textrm{decay}}$ & decay & rate of exponential decay \\
    & decay-step & decay step & number iterations (steps) until the next decay is applied \\
    \hline
    \end{tblr}
    \label{tab:overview-param}
\end{table}